\documentclass[aps,preprintnumbers,showkeys,showpacs,nofootinbib,byrevtex,superscriptaddress,fleqn]{revtex4-1}   
\usepackage{amsmath,amsfonts,amssymb,amscd,amsxtra,amsthm}
\usepackage{graphicx,slashed}
\usepackage{bm}

\usepackage[normalem]{ulem} 
\usepackage[dvips]{color} 
\renewcommand\sout{\bgroup \color{red} \ULdepth=-.5ex \ULset}

\begin{document}   
\preprint{INHA-NTG-16/2011}
\preprint{KIAS-P11064}
\title{Contribution of higher nucleon resonances to $K^*\Lambda$
  photoproduction}       
\author{Sang-Ho Kim}
\email[E-mail: ]{sanghokim@ihha.edu}
\affiliation{Department of Physics, Inha University, Incheon 402-751, Korea}  
\author{Seung-il Nam}
\email[E-mail: ]{sinam@kias.re.kr}
\affiliation{School of Physics, Korea Institute for Advanced Study,
  Seoul 130-722, Korea} 
\author{Yongseok Oh}
\email[E-mail: ]{yohphy@knu.ac.kr}
\affiliation{School of Physics, Korea Institute for Advanced Study,
  Seoul 130-722,  Korea} 
\affiliation{Department of Physics, Kyungpook National University,
  Daegu 702-701, Korea}  
\author{Hyun-Chul Kim}
\email[E-mail: ]{hchkim@inha.ac.kr}
\affiliation{Department of Physics, Inha University, Incheon 402-751,
  Korea}
\affiliation{School of Physics, Korea Institute for Advanced Study,
  Seoul 130-722, Korea} 

\date{\today}
\begin{abstract}
We investigate $K^*\Lambda(1116)$ photoproduction off the proton and
neutron targets in the effective Lagrangian method in the tree-level
Born approximation. In addition to the $K$, $K^*$, and $\kappa$
exchanges which are the main contributions to the reaction, we
consider the contributions from the higher nucleon 
resonances $D_{13}(2080)$ and the $D_{15}(2200)$. We find that the
$D_{13}(2080)$ plays a crucial role in explaining the enhancement of
the near-threshold production rate, which results in a 
good agreement with the experimental data for the proton targets,  
while the contribution of the $D_{15}(2200)$ is rather small. A
similar conclusion is drawn for $K^*$ photoproduction off the neutron 
targets. In addition to the energy and angular dependence of the cross
sections, we present the predictions on the photon-beam asymmetry.  
\end{abstract} 
\pacs{13.60.Le, 14.20.Gk}
\keywords{$K^*$ photoproduction, effective Lagrangian method, nucleon
resonance, polarization observables}   
\maketitle
\section{Introduction}
Strange meson and baryon photoproductions provide a very useful tool  
in studying the properties of strange hadrons and, in particular,
in examining the structure of non-strange baryon resonances that
cannot be seen in the non-strange $\pi N$ channel. This was shown
manifestly in the investigation of kaon photoproduction off the
nucleon targets, i.e., $\gamma N \to K \Lambda(1116)$, which has been
studied intensively both theoretically and
experimentally~\cite{CLAS05c,MAMI11,HKT10,TTWF07,WBDE06,JRDV02,
Mart11,YCK11,DIVR10}. These works show that nucleon resonances play a
crucial role in understanding the production mechanisms of the reaction
in the resonance region~\cite{JRDV02,Mart11}.  

The recent upgrade plans of the photon-beam facilities such as the
LEPS2 project at SPring-8~\cite{LEPS07b} and the CLAS12 project at
Thomas Jefferson National Accelerator Facility (TJNAF)~\cite{CLAS10a}
will enable us to study the photoproduction processes of heavier
strange-mesons with an unprecedented accuracy. This will enrich our
understanding of strange hadrons. In particular, the production of
strange vector mesons can be investigated in detail. The experimental
studies on the exclusive production of the $K^*$ vector mesons from
photon-nucleon scattering have been reported only 
recently~\cite{CLAS06e,CLAS07a,CBELSA/TAPS-08}%
\footnote{The numerical error of a factor of $3/2$ in the data reported in
  Refs.~\cite{CLAS06e,CLAS07a} was pointed out in
  Ref.~\cite{CBELSA/TAPS-08} and was corrected later. See
  Ref.~\cite{CLAS07a}.}. 
These experiments have measured mostly the total and the differential 
cross sections and have shown that although the cross sections
of $K^*$ photoproduction are small compared to those  
of kaon photoproduction, their values are non-negligible. It
indicates that production reactions with heavier strange hadrons
should be considered when analyzing higher nucleon resonances, which
are still less known theoretically as well as experimentally (see,
for example, the Particle Data Group (PDG) compilation for the $N^*$
resonances~\cite{PDG10}). We expect that the upgrades of the
experimental facilities can offer a chance to study various physical
quantities of $K^*$ photoproduction in detail so that one can have
enough data for unraveling the excited spectra of the non-strange
baryon resonances. Theoretically, $K^*$ photoproduction was studied in
a quark model~\cite{ZAB01} and in the effective Lagrangian 
approach~\cite{OK06a,OK06b}. In particular,  
Refs.~\cite{OK06a,OK06b} have emphasized the role of the $t$-channel
scalar $\kappa$ meson exchange and have suggested its experimental
test. 

Recently, new experimental data were announced by the CLAS Collaboration
at TJNAF~\cite{HKT10,CLASKSTAR} for the angular distribution of the
cross sections for the reaction of $\gamma p\to K^{*+}\Lambda$. When
compared with previous theoretical
calculations~\cite{OK06a,OK06b,KNOK11}, the new data show that  
the prediction of the model of Refs.~\cite{OK06a,OK06b} underestimates
the cross sections near the threshold, which is in the $N^*$ resonance
region. Therefore, it is legitimate to reconsider the theoretical
model of Refs.~\cite{OK06a,OK06b} with higher nucleon resonances being
taken into account. 

In the present paper, we investigate $K^*\Lambda$ photoproduction off
the nucleon targets, i.e., $\gamma p\to K^{*+}\Lambda$ and $\gamma
n\to K^{*0}\Lambda$. Employing the effective Lagrangian method in
the tree-level Born approximation, we improve the model of
Refs.~\cite{OK06a,OK06b} by including the contributions of nucleon
resonances. Because of isospin conservation, intermediate $\Delta$
resonances are not allowed and nucleon resonances can only contribute
to this process. Therefore, it is of great significance to extract
carefully parameters of the $N^*$ resonances such as their coupling
strengths to the $K^*$ and the $\Lambda(1116)$. More reliable resonance
parameters might be obtained by full partial-wave analyses. However, 
since the threshold energy of $K^*\Lambda$ production is about 2~GeV,
it is reasonable to start with only the $N(2080,3/2^-)$ and the   
$N(2200,5/2^-)$ as relevant nucleon resonances for this
process near the threshold, other resonances being neglected. We
will show that these two resonances are enough to address the
important role of nucleon resonances in this reaction.   
Following the model of Ref.~\cite{OK06b}, we also take into account
other production mechanisms that include the nucleon $(N)$ and hyperon
$(\Lambda,\Sigma,\Sigma^*)$ pole contributions in the $s$ and $u$
channels, respectively, and the exchanges of strange mesons
$(\kappa,K,K^*)$ in the $t$ channel. A contact-term contribution is
considered as well in order to fulfill the current conservation
of the electromagnetic interactions.  

This paper is organized as follows. In Sec.~II, we briefly introduce
the framework of our model. The effective Lagrangians for the
resonance interactions are given explicitly as well as those for the
background production mechanisms. All the parameters used in the
present calculation are given explicitly.
We compare our results of the cross sections with the experimental
data for the reaction of $\gamma p \to K^*\Lambda$ in Sect.~III. 
The predictions on the photon beam asymmetries are then discussed. 
We also discuss the dependence of these physical quantities on
isospin. Section~IV summarizes and draws conclusions.

\section{Formalism}
\begin{figure*}[t]
\includegraphics[width=12cm]{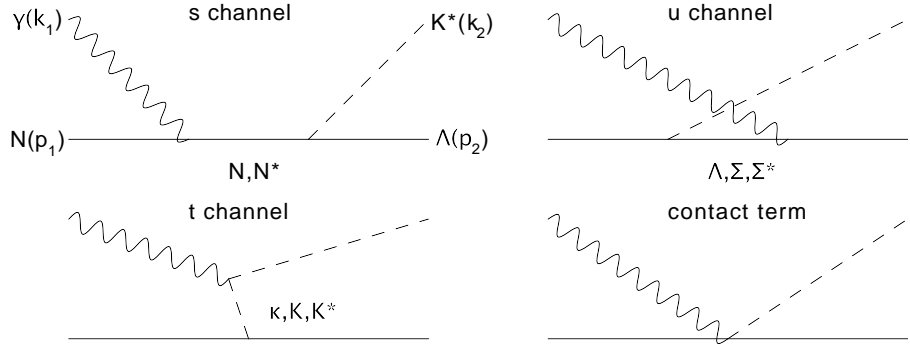}
\caption{Feynman diagrams for the $\gamma N\to K^*\Lambda$ reaction.}       
\label{FIG1}
\end{figure*}

In the present work, we employ the effective Lagrangian method at the
tree-level Born approximation. The relevant Feynman diagrams for the
reaction mechanisms of $\gamma N \to K^* \Lambda$ are shown in
Fig.~\ref{FIG1}, which include the nucleon and nucleon
resonance poles in the $s$-channel, the $K^*$, $K$, and $\kappa$ meson
exchanges in the $t$-channel, and $\Lambda$, $\Sigma$, and
$\Sigma^*(1385,3/2^+)$ hyperon terms in the $u$-channel. The contact
term is also taken into account to preserve the gauge invariance.

The effective Lagrangians for the background production mechanisms are
essentially the same as those used in Refs.~\cite{OK06a,OK06b}. As for 
the photon interaction Lagrangians of strange mesons, we have 
\begin{eqnarray}
\label{eq:LAGEM}
\mathcal L_{\gamma K^* K^*} &=& 
-ie_{K^*}^{} A^\mu \left( K^{*\nu} K_{\mu\nu}^{*\dagger} -
  K_{\mu\nu}^{*} K^{*\dagger\nu} \right), 
\nonumber \\ 
\mathcal L_{\gamma K^*K} &=&
g_{\gamma KK^*}^{} \varepsilon^{\mu\nu\alpha\beta} \left( \partial_\mu
  A_\nu \right) 
\left(\partial_\alpha K_\beta^* \right) K+\mathrm{H.c.},
\nonumber \\
\mathcal{L}_{\gamma K^* \kappa} &=&
g_{\gamma K^* \kappa}^{} A^{\mu\nu} \kappa K^*_{\mu\nu} + \mathrm{H.c.},
\label{eq1}
\end{eqnarray}
where $A_\mu$, $K^*_\mu$, $K$, and $\kappa$ denote the photon, the
$K^*(892,1^-)$ vector meson, the $K(495,0^-)$ pseudoscalar meson, and
the $\kappa(800,0^+)$ scalar meson, respectively, and
$K^*_{\mu\nu}=\partial_\mu K^*_\nu-\partial_\nu K^*_\mu$.  
The electric charge of the $K^*$ vector meson is denoted by
$e_{K^*}^{}$. We take the values of $g_{\gamma K^*K}$ from the
experimental data given by the PDG~\cite{PDG10},
which leads to $g^\mathrm{charged}_{\gamma K^*K}=0.254$~GeV$^{-1}$ and  
$g^\mathrm{neutral}_{\gamma K^*K}=-0.388$~GeV$^{-1}$, whereas we use the
vector-meson dominance~\cite{BHS02} to determine the values
of $g_{\gamma K^* \kappa}^{}$: $g^\mathrm{charged}_{\gamma
  K^*\kappa}=0.12\,e$~GeV$^{-1}$ and $g^\mathrm{neutral}_{\gamma
  K^*\kappa}=0.24\,e $~GeV$^{-1}$ with the unit electric charge
  $e$.  

The Lagrangians for the electromagnetic interactions of baryons are
given as 
\begin{eqnarray}
\mathcal L_{\gamma NN}&=&
-\bar{N }\left [e_N\slashed{A} -\frac{e \kappa_N^{}}{2M_N}
\sigma_{\mu\nu}\partial^\nu A^\mu \right ] N,
\nonumber \\
\mathcal L_{\gamma\Lambda\Lambda}&=&
\frac{e \kappa_\Lambda^{}}{2M_N}
\bar{\Lambda}\sigma_{\mu\nu}\partial^\nu A^\mu \Lambda,
\nonumber \\
\mathcal L_{\gamma\Lambda\Sigma}&=&
\frac{e \mu_{\Sigma\Lambda}^{}}{2M_N} \bar{\Sigma}\sigma_{\mu\nu}
\partial^\nu A^\mu \Lambda +\mathrm{H.c.},
\nonumber \\
\mathcal L_{\gamma\Lambda\Sigma^*}&=&
-\frac{ie }{2M_N}\left [g^V_{\gamma\Lambda\Sigma^*}\bar{\Lambda}
\gamma_\nu-\frac{ig^T_{\gamma\Lambda\Sigma^*}}
{2M_N}(\partial_\nu\bar{\Lambda})\right ]
\gamma_5\Sigma_\mu^*F^{\mu\nu} + \mathrm{H.c.},
\label{eq:GBB}
\end{eqnarray}
where $N$, $\Lambda$, $\Sigma$, and $\Sigma^*$ stand for the nucleon, 
$\Lambda(1116)$, $\Sigma(1192)$, and $\Sigma^*(1385,3/2^+)$ hyperon
fields, respectively, and $M_h$ denotes the mass of hadron $h$. The
baryon fields with spin $s \ge 3/2$ are described by the
Rarita-Schwinger vector-spinor formalism~\cite{RS41,Read73}. Here,
$\kappa_B^{}$ is the anomalous magnetic moment of baryon $B$ and
$\mu_{\Lambda\Sigma}$  is the transition magnetic moment between the
$\Lambda(1116)$ and the $\Sigma(1192)$. Their PDG values are given in
Table~\ref{TABLE1}. The electromagnetic (EM) coupling with the
spin-3/2 hyperon $\Sigma^*$ has two independent terms as shown in the last
line of Eq.~(\ref{eq:GBB}). These coupling constants are determined by
the experimental data for the decay width
$\Gamma_{\Sigma^*\to\gamma\Lambda}$~\cite{PDG10}, which leads to
$(g^V_{\gamma\Lambda\Sigma^*},g^T_{\gamma\Lambda\Sigma^*})=(3.78,3.18)$.  
The EM couplings in Eqs.~(\ref{eq1}) and (\ref{eq:GBB}) are summarized
in Table~\ref{TABLE1}. 
\begin{table}[t]
\begin{tabular}{cc|cc|cccc|cc} \hline\hline
$g^\mathrm{charged}_{\gamma K^*K}$
&$g^\mathrm{neutral}_{\gamma K^*K}$
&$g^\mathrm{charged}_{\gamma K^*\kappa}$
&$g^\mathrm{neutral}_{\gamma K^*\kappa}$
&$\kappa_p$
&$\kappa_n$
&$\kappa_\Lambda$
&$\mu_{\Lambda\Sigma}$
&$g^V_{\gamma\Lambda\Sigma^*}$
&$g^T_{\gamma\Lambda\Sigma^*}$\\
\hline
$0.254/$GeV
&$-0.388/$GeV
&$0.12\,e /$GeV
&$0.24\,e /$GeV
&$1.79\mu_N$
&$-1.91\mu_N$
&$-0.61\mu_N$
&$(1.61\pm0.08)\mu_N$
&$3.78$
&$3.18$\\
\hline\hline
\end{tabular}
\caption{Electromagnetic coupling constants for the interactions in
  Eqs.~(\ref{eq1}) and (\ref{eq:GBB}).} 
\label{TABLE1}
\end{table}

The effective Lagrangians for the meson-baryon interactions are
\begin{eqnarray}
\mathcal L_{K^* N\Lambda} &=& -g_{K^* N \Lambda}^{} \bar{N} \Lambda
\left[\slashed{K}^*- \frac{\kappa_{K^* N\Lambda}^{} }{2M_N}
\sigma_{\mu\nu}\partial^\nu K^{*\mu}\right] + \mbox{H.c.}, 
\nonumber \\
\mathcal L_{K^* N\Sigma} &=& -g_{K^* N \Sigma}^{} \bar{N} \Sigma
\left[\slashed{K}^*- \frac{\kappa_{K^* N\Sigma}^{} }{2M_N}
\sigma_{\mu\nu}\partial^\nu K^{*\mu}\right] + \mbox{H.c.}, 
\nonumber \\
\mathcal L_{KN\Lambda} &=&
-ig_{KN\Lambda}^{} \bar{N} \gamma_5 \Lambda K + \mathrm{H.c.},       
\nonumber \\
\mathcal{L}_{\kappa N\Lambda} &=&
-g_{\kappa N\Lambda}^{} \bar{N}\kappa\Lambda + \mathrm{H.c.},     
\nonumber \\
\mathcal L_{K^* N\Sigma^*}
&=& -\frac{if^{(1)}_{K^* N\Sigma^*}}{2M_{K^*}}\bar{K^*}_{\mu\nu}\bar{\Sigma}^{*\mu}
\gamma^\nu \gamma_5 N -\frac{f^{(2)}_{K^* N\Sigma^*}}
{4M_{K^*}^2}\bar{K}^*_{\mu\nu}
\bar{\Sigma}^{*\mu} \gamma_5 \partial^\nu N 
+\frac{f^{(3)}_{K^* N\Sigma^*}}{4M_{K^*}^2}
\partial^\nu\bar{K}^*_{\mu\nu}\bar{\Sigma}^{*\mu} \gamma_5 N
+\mathrm{H.c.},
\label{eq:MBB}
\end{eqnarray}
where $\Sigma = \bm{\tau} \cdot \bm{\Sigma}$ and $\Sigma^*_\mu =
\bm{\tau} \cdot \bm{\Sigma}^*_\mu$. The strong coupling constants can
be estimated by the Nijmegen soft-core model (NSC97a)~\cite{RSY99},  
and the corresponding values are presented in Table.~\ref{TABLE2}. 
Considering the minimally possible Lorentz structure for the vector meson 
coupling to the $\Sigma^*$, we can write the interaction Lagrangian
with the three form factors as shown in the last line of  
Eq.~(\ref{eq:MBB}), which is similar to the case of
$\mathcal{L}_{\gamma\Lambda\Sigma^*}$. From the flavor SU(3)
symmetry~\cite{OK04}, the value for $f^{(1)}_{K^* N\Sigma^*}$ can
be estimated. Because of the lack of experimental and theoretical
information on $f^{(2,3)}_{K^* N\Sigma^*}$ we do not consider these
terms in the present work, following Ref.~\cite{OK04}.  
Finally, we have the contact interaction given by 
\begin{equation}
\mathcal L_{\gamma K^* N\Lambda}=
-\frac{ie_{K^*}^{} g_{K^*N\Lambda}^{} \kappa_{K^* N\Lambda}^{} }{2M_N}
\bar{\Lambda}\sigma^{\mu\nu} A_\nu K^{*}_\mu N+\mathrm{H.c.},
\label{eq:CONT}
\end{equation}
which is obtained by the minimal gauge substitution $\partial_\mu\to ie_{K^*}A_\mu$
to the $K^*N\Lambda$ interaction. 

\begin{table*}[t]
\begin{tabular}{ccccccc} \hline\hline
$g_{K^*N\Lambda}$&
$g_{K^*N\Sigma}$&
$g_{KN\Lambda}$&
$g_{\kappa N\Lambda}$&
$f^{(1)}_{K^*N\Sigma^*}$&
$f^{(2)}_{K^*N\Sigma^*}$&
$f^{(3)}_{K^*N\Sigma^*}$\\
\hline
$-4.26$&
$-2.46$&
$-13.24$&
$-8.3$&
$5.21$&
$0$&
$0$\\ \hline\hline
\end{tabular}
\caption{Strong coupling constants for the meson-baryon interactions
  in Eq.~(\ref{eq:MBB}).} 
\label{TABLE2}
\end{table*}

Considering all the ingredients discussed so far, we obtain the
scattering amplitude of each channel as given in Fig.~\ref{FIG1}:
\begin{eqnarray}
\label{BRONTERM}
\mathcal{M}_{t(K^*)} &=& \frac{e_{K^*}^{} g_{K^* N\Lambda}^{}
}{t-M_{K^*}^2} \varepsilon_\nu^* \bar{u}_\Lambda^{}
\left [ 2k_2^\mu g^{\nu\alpha} - k_2^\alpha g^{\mu\nu} + k_1^\nu
  g^{\mu\alpha} \right ] \left [ g_{\alpha\beta} - \frac{q_\alpha
    q_\beta}{M_{K^*}{}^2}\right ]  \left [ \gamma^\beta -
  \frac{i\kappa_{K^* N\Lambda}^{} }{2M_N} \sigma^{\beta \delta}
  q_\delta \right ] u_N^{} \epsilon_\mu, 
\nonumber \\
\mathcal{M}_{t(K)} &=& \frac{i g_{\gamma KK^*}^{} g_{KN\Lambda}^{}
}{t-M_K^2} \varepsilon_\nu^* 
\bar{u}_\Lambda^{} \epsilon^{\mu\nu\alpha\beta} k_{1\alpha} k_{2\beta}
\gamma_5^{} u_N^{} \epsilon_\mu,
\nonumber \\
\mathcal{M}_{t(\kappa)} &=& \frac{-2g_{\gamma K^* \kappa}^{} g_{\kappa
    N\Lambda}^{}} {t-(M_\kappa-i\Gamma_\kappa/2)^2} \varepsilon_\nu^* 
\bar{u}_\Lambda^{} (k_1 \cdot k_2 g^{\mu\nu}-k_1^\nu k_2^\mu) u_N^{}
\epsilon_\mu,
\nonumber \\
\mathcal{M}_{s(N) } &=& \frac{g_{K^* N\Lambda}^{} }{s-M_N^2}
\varepsilon_\nu^*\bar{u}_\Lambda^{}       
\left [ \gamma^\beta - \frac{i\kappa_{K^* N\Lambda}^{} }{2M_N}
{\sigma^{\beta \delta}}q_\delta \right ]
(\slashed{k}_1+\slashed{p}_1+M_N)                         
\left [ e_N^{} \gamma^\mu +\frac{ie \kappa_N^{} }{2M_N}
\sigma^{\mu\nu} k_\nu \right ]u_N^{} \epsilon_{\mu},
\nonumber \\
\mathcal M_{u(\Lambda)} &=& \frac{ie \kappa_\Lambda^{} }{2M_N}
\frac{g_{K^*N^*\Lambda}^{} }{u-M_\Lambda^2}
\varepsilon_\nu^* \bar{u}_\Lambda^{} \sigma^{\mu\nu} k_{2\nu}
(\slashed{p}_1-\slashed{k}_2+M_\Lambda) u_N^{} \epsilon_{\mu},            
\nonumber \\
\mathcal M_{u(\Sigma)} &=& \eta_\Sigma^{} \frac{ie \mu_{\Sigma\Lambda}^{} }{2M_N}
\frac{g_{K^*N^*\Lambda}^{} }{u-M_\Sigma^2}
\varepsilon_\nu^* \bar{u}_\Lambda^{} \sigma^{\mu\nu} k_{2\nu}
(\slashed{p}_1-\slashed{k}_2+M_\Sigma) u_N^{} \epsilon_\mu,
\nonumber \\
\mathcal{M}_{u(\Sigma^*)} &=& \eta_{\Sigma^*}^{} \varepsilon_\nu^*
\bar{u}_\Lambda^{} 
(k_1^\beta g^{\mu \delta}-k_1^\delta g^{\mu\beta})
\left [\frac{g_1^{}}{2M_N} \gamma_\delta+\frac{g_2^{}}{4M_N^2}
p_{2\delta}\right ] \gamma_5^{} 
\Delta_{\beta\alpha} \frac{ef_1^{}}{2M_{K^*}} \gamma_\lambda^{}
\gamma_5^{} (k_2^\alpha g^{\nu \lambda} - k_2^\lambda g^{\nu\alpha}) 
u_N^{} \epsilon_\mu,
\nonumber \\
\mathcal{M}_\mathrm{contact}
&=&-\frac{ie_{K^*}^{} g_{K^*N\Lambda}^{} \kappa_{K^*N\Lambda}^{}}{2M_N}
\varepsilon_\nu^* \bar{u}_\Lambda^{} \sigma^{\mu\nu} u_N^{}
\epsilon_\mu, 
\end{eqnarray}
where $\varepsilon_\mu^*$ and $\epsilon_\mu$ indicate the polarization
vectors for the outgoing $K^*$ and incoming photon, respectively, and
$(s,u,t)$ denote the usual Mandelstam variables.  
The reaction with the proton (neutron) targets corresponds to 
$\eta_{\Sigma}^{}=+1$ ($-1$). The spin-3/2 Rarita-Schwinger spin
projection is expressed as 
\begin{equation}
\Delta_{\beta\alpha}=\frac{\slashed{p}+M_{\Sigma^*}}
{{p^2-M^2_{\Sigma^*}}}\left [ - g_{\beta\alpha}^{}
+\frac{1}{3}\gamma_\beta^{} \gamma_\alpha^{} +
\frac{1}{3M_{\Sigma^*}}(\gamma_\beta^{} p_\alpha^{} - \gamma_\alpha^{}
p_\beta^{}) +\frac{2}{3M^2_{\Sigma^*}}p_\beta^{} p_\alpha^{} \right
]. 
\end{equation}

In addition to the production mechanisms discussed above, we now
include the nucleon resonance contributions. Here, we consider the
$D_{13}(2080)$ and $D_{15}(2200)$, which lie close to the
threshold of $K^*$ photoproduction. This can be established by the
following effective Lagrangians: 
\begin{eqnarray}
\label{eq:LAGEMRE}
\mathcal{L}_{\gamma  ND_{13}}
&=& -\frac{ieh_{1D_{13}}^{} }{2M_N} \bar N  \Gamma_\nu^- F^{\mu\nu}
R_\mu -\frac{eh_{2D_{13}}^{} }{(2M_N)^2} \partial_\nu \bar N \Gamma^-
F^{\mu\nu} R_\mu  + \mathrm{H.c.}\,,                       
\nonumber \\
\mathcal{L}_{\gamma ND_{15}} 
&=& \frac{eh_{1D_{15}}^{} }{(2M_N)^2} \bar N
\Gamma_\nu^+ \partial^\alpha  F^{\mu\nu} R_{\mu\alpha}
-\frac{ieh_{2D_{15}}^{} }{(2M_N)^3} \partial_\nu \bar N
\Gamma^+ \partial^\alpha  F^{\mu\nu} R_{\mu\alpha} + \mathrm{H.c.},  
\end{eqnarray}
where $R$ stands for the field for the nucleon resonances with a
certain spin and parity, and  
\begin{equation}
\label{eq:GAMMASPEM}
\Gamma_\mu^{(\pm)} = \left(
\begin{array}{c}
\gamma_\mu^{} \gamma_5^{} \\ \gamma_\mu^{}
\end{array} \right),
\qquad
\Gamma^{(\pm)} = \left(
\begin{array}{c} 
\gamma_5^{} \\ \mathbf{1}
\end{array} \right).
\end{equation}
The coupling constants are determined by using the experimental data
for the  helicity amplitudes~\cite{PDG10} and the quark model
predictions of Ref.~\cite{Caps92}, which gives $h_{1D_{13}}^{} =
0.608$, $h_{2D_{13}}^{} = -0.620$, $h_{1D_{15}}^{} = 0.123$, and
$h_{2D_{15}}^{} =0.011$. The effective Lagrangians for the relevant
strong interactions are  
\begin{eqnarray}
\label{eq:STRONG}
\mathcal{L}_{K^* D_{13} \Lambda }
&=& -\frac{ig_{1D_{13}}^{} }{2M_N} \bar \Lambda \Gamma_\nu^- 
K^{*\mu\nu} R_\mu -\frac{g_{2D_{13}}^{} }{(2M_N)^2} \partial_\nu \bar
\Lambda  \Gamma^- K^{*\mu\nu} R_\mu                                          
+\frac{g_{3D_{13}}^{} }{(2M_N)^2} \bar \Lambda \Gamma^- 
\partial_\nu  K^{*\mu\nu} R_\mu + \mathrm{H.c.} \,,     
\nonumber \\
\mathcal{L}_{K^* D_{15}\Lambda }\,,
&=&  \frac{g_{1D_{15}}^{} }{(2M_N)^2} \bar \Lambda \Gamma_\nu^+
 \partial^\alpha K^{*\mu\nu} R_{\mu\alpha}
 -\frac{ig_{2D_{15}}^{} }{(2M_N)^3} \partial_\nu \bar \Lambda \Gamma^+
\partial^\alpha K^{*\mu\nu} R_{\mu\alpha} + \frac{ig_{3D_{15}}^{} }{(2M_N)^3} 
\bar \Lambda \Gamma^+ \partial^\alpha \partial_\nu K^{*\mu\nu}
R_{\mu\alpha} + \mathrm{H.c.} . 
\end{eqnarray}
The corresponding scattering amplitudes can be obtained as
\begin{eqnarray}
{\cal M}_{s(R)} \left( {\frac{3}{2}^\pm} \right) 
&& = \frac{1}{s-M_R{}^2} \bar \epsilon^*_\nu u_\Lambda^{}
\left[ \frac{g_1^{}}{2M_N}\Gamma_\sigma^{(\pm)} +
  \frac{g_2^{}}{(2M_N)^2} p_{2 \sigma}^{} \Gamma^{(\pm)} 
- \frac{g_3^{}}{(2M_N)^2} k_{2 \sigma}^{} \Gamma^{(\pm)} \right]                                                             
( k_2^\beta g^{\nu \sigma} - k_2^\sigma g^{\nu \beta})   \nonumber \\
&& \mbox{} \times ( \slashed{k}_1^{} + \slashed{p}_1^{} + M_R ) \Delta_{\beta\alpha}(R,k_1^{}+p_1^{})                                                       
\left[ \frac{e\mu_R^{}}{2M_N} \Gamma_\delta^{(\pm)} \,\mp\, 
\frac{e\bar \mu_R^{}}{(2M_N)^2} \Gamma^{(\pm)} p_{1 \delta}^{} \right]
(k_1^\alpha g^{\mu \delta} - k_1^\delta g^{\alpha\mu}) u_N^{}
\epsilon_\mu   \nonumber                            
\end{eqnarray}                                                                     
\begin{eqnarray}
{\cal M}_{s(R)} \left( {\frac{5}{2}^\pm} \right) 
&& = \frac{1}{s-M_R{}^2} \epsilon^*_\nu \bar u_\Lambda^{}
\left[ \frac{g_1^{}}{(2M_N)^2}\Gamma_\sigma^{(\mp)} +
  \frac{g_2^{}}{(2M_N)^3} p_{2 \sigma}^{} \Gamma^{(\mp)} 
- \frac{g_3^{}}{(2M_N)^3} k_{2 \sigma}^{} \Gamma^{(\mp)} \right]                                                        
k_2^\delta ( k_2^\rho g^{\nu \sigma} - k_2^\sigma g^{\nu \rho})   \nonumber \\
&& \mbox{} \times ( \slashed{k}_1 + \slashed{p}_1 + M_R ) \Delta_{\rho \delta ; \alpha
  \beta}(R,k_1^{}+p_1^{})                             
\left[ \frac{e\mu_R^{}}{(2M_N)^2} \Gamma^{(\mp)}_j \,\pm\, 
       \frac{e\bar \mu_R^{}}{(2M_N)^3} \Gamma^{(\mp)} p_{1 j}^{} \right]
k_1^\beta ( k_1^\alpha g^{\mu j} - k_1^j g^{\alpha \mu} ) u_N^{}
\epsilon_\mu\,, \cr
&&
\end{eqnarray}
where the spin-5/2 Rarita-Schwinger spin projection is given by 
\begin{eqnarray}
& &\Delta_{\rho\delta ; \alpha\beta}(p,M)                                 
=\frac{1}{2}(\bar g_{\rho\alpha} \bar g_{\delta\beta} + \bar
g_{\rho\beta} \bar g_{\delta\alpha})  - \frac{1}{5} \bar
g_{\rho\delta} \bar g_{\alpha\beta} - \frac{1}{10}  
( \bar \gamma_\rho \bar \gamma_\alpha \bar g_{\delta\beta} + \bar
\gamma_\rho \bar \gamma_\beta \bar g_{\delta\alpha} 
 +\bar \gamma_\delta \bar \gamma_\alpha \bar g_{\rho\beta} + \bar
 \gamma_\delta \bar \gamma_\beta \bar g_{\rho\alpha} )\,,
\end{eqnarray}
where
\begin{equation}
\bar g_{\alpha\beta} = g_{\alpha\beta} - \frac{p_\alpha
    p_\beta}{M^2}, \qquad \bar \gamma_\alpha = \gamma_\alpha - \frac{p_\alpha
  }{M^2} \not p . 
\end{equation}
The strong coupling constants in Eq.~(\ref{eq:STRONG}) can be
determined from the theoretical estimations for the partial-wave decay
amplitudes, 
\begin{equation}
\label{eq:GGG}
\Gamma_{R\to K^*\Lambda}=\sum_\ell|G(\ell)|^2,
\end{equation}
where $\Gamma_{R\to K^*\Lambda}$ is the decay width of $R \to K^* \Lambda$.
The values for the partial-wave coupling strengths $G(\ell)$ can be found,
e.g., in Ref.~\cite{CR98b}. Since the purpose of the present work is to
investigate the role of nucleon resonances near the threshold in the
Born approximation, it must be a good approximation to take into account  
only the low partial-wave contributions. Hence, we consider only the
$g_1^{}$ terms in Eq.~(\ref{eq:STRONG}), employing only the lowest  
partial-wave contribution for $G(\ell)$. 
Then by using Eq.~(\ref{eq:GGG}) and the prediction of Ref.~\cite{CR98b}, 
we obtain $|g_{1D_{13}}|=1.59$ and $|g_{1D_{15}}|=1.03$. 
The signs of these strong coupling constants are determined by fitting
the experimental data. We list all the parameters of the resonances in
Table~\ref{TABLE3}. 

\begin{table}[t]
\begin{tabular}{cccc|cccc|ccc|ccc} \hline\hline
$h^{p}_{1D_{13}}$&
$h^{p}_{2D_{13}}$&
$h^{p}_{1D_{15}}$&
$h^{p}_{2D_{15}}$&
$h^{n}_{1D_{13}}$&
$h^{n}_{2D_{13}}$&
$h^{n}_{1D_{15}}$&
$h^{n}_{2D_{15}}$&
$g_{1D_{13}}^{}$&
$g_{2D_{13}}^{}$&
$g_{3D_{13}}^{}$&
$g_{1D_{15}}^{}$&
$g_{2D_{15}}^{}$&
$g_{3D_{15}}^{}$\\
\hline
$0.608$&
$-0.620$&
$0.123$&
$0.011$&
$-0.770$&
$0.531$&
$-0.842$&
$-0.782$&
$-1.59$&
$0$&
$0$&
$1.03$&
$0$&
$0$\\ \hline\hline
\end{tabular}
\caption{Coupling constants for the resonances in
  Eqs.~(\ref{eq:LAGEMRE}) and (\ref{eq:STRONG}).} 
\label{TABLE3}
\end{table}

In the low-energy region where we are interested in, the hadron can
not be simply viewed as a point-like object but spatially extended
one, characterized by the phenomenological form factors in
general. Hence, it is necessary to include the EM and hadronic form
factors that represent the structure of hadrons probed by the photon
and mesons, which violates gauge invariance of the amplitude.  
Thus, following the prescription described in
Refs.~\cite{HBMF98a,DW01a,HNK06}, we write the reaction amplitudes as 
\begin{equation}
\mathcal{M}_\mathrm{proton}=
(\mathcal{M}_{K^*}+\mathcal{M}_p
+\mathcal{M}_\mathrm{contact})F_\mathrm{common}^2 
+ \mathcal{M}_KF_K^2 + \mathcal{M}_\kappa F_\kappa^2 
+ \mathcal{M}_\Lambda F_\Lambda^2 
+ \mathcal{M}_\Sigma F_\Sigma^2 
+ \mathcal{M}_{\Sigma^*} F_{\Sigma^*}^2
+ \mathcal{M}_{N^*}F_{N^*}^2   
\end{equation}
for $\gamma p \to K^{*+}\Lambda$ and
\begin{equation}
\mathcal{M}_\mathrm{neutron} = \mathcal{M}_KF_K^2
 + \mathcal{M}_\kappa F_\kappa^2 
 + \mathcal{M}_n F_n^2
 + \mathcal{M}_\Lambda F_\Lambda^2 
 + \mathcal{M}_\Sigma F_\Sigma^2 
 + \mathcal{M}_{\Sigma^*}F_{\Sigma^*}^2 
 + \mathcal{M}_{N^*}F_{N^*}^2
\end{equation}
for $\gamma n \to K^{*0}\Lambda$.
The form factors are defined generically as
\begin{equation}
\label{eq:FF}
F_\mathrm{common}
=F_pF_{K*}-F_p-F_{K*}, \qquad
F_\Phi=\frac{\Lambda^2_{\Phi}-M^2_\Phi}
{\Lambda^2_\Phi-p^2},
\qquad
F_B=\frac{\Lambda^4_B}{\Lambda^4_B+(p^2-M^2_B)^2}\,,
\end{equation}
where $p$ denotes the off-shell momentum of the relevant hadron. For
the mesonic $(\Phi=K^*,K,\kappa)$ and baryonic
$(B=N,N^*,\Lambda,\Sigma,\Sigma^*,R)$ vertices, we consider different
types of the form factors with the cutoff parameters $\Lambda_\Phi$
and $\Lambda_B$. Those used in the present work are summarized in
Table.~\ref{TABLE4}, which are determined to reproduce the available
cross section data for $K^*$ photoproduction.  

\begin{table}[t]
\begin{tabular}{ccccccccc} \hline\hline
$\Lambda_{K*}$&
$\Lambda_{K}$&
$\Lambda_{\kappa}$&
$\Lambda_{N}$&
$\Lambda_{\Lambda}$&
$\Lambda_{\Sigma}$&
$\Lambda_{\Sigma^*}$&
$\Lambda_{D_{13}}$&
$\Lambda_{D_{15}}$\\
\hline
$0.9$ GeV&
$1.25$ GeV&
$1.25$ GeV&
$0.9$ GeV&
$0.9$ GeV&
$0.9$ GeV&
$0.9$ GeV&
$1.2$ GeV&
$1.2$ GeV\\ \hline \hline
\end{tabular}
\caption{Cutoff parameters in Eq.~(\ref{eq:FF}) for each channel.} 
\label{TABLE4}
\end{table}

\section{Numerical results and Discussions}
\begin{figure}[t]
\includegraphics[width=7.5cm]{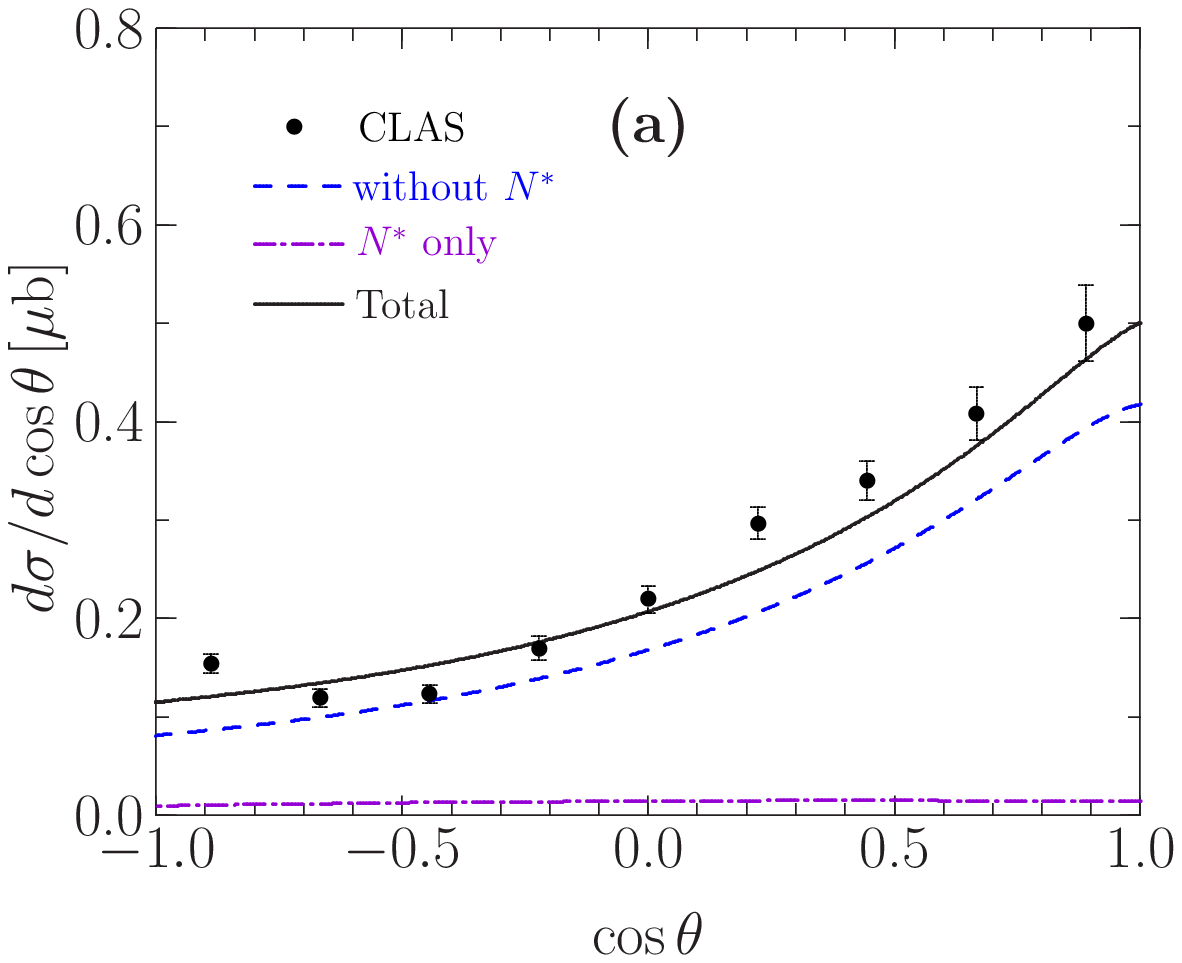} \qquad
\includegraphics[width=7.5cm]{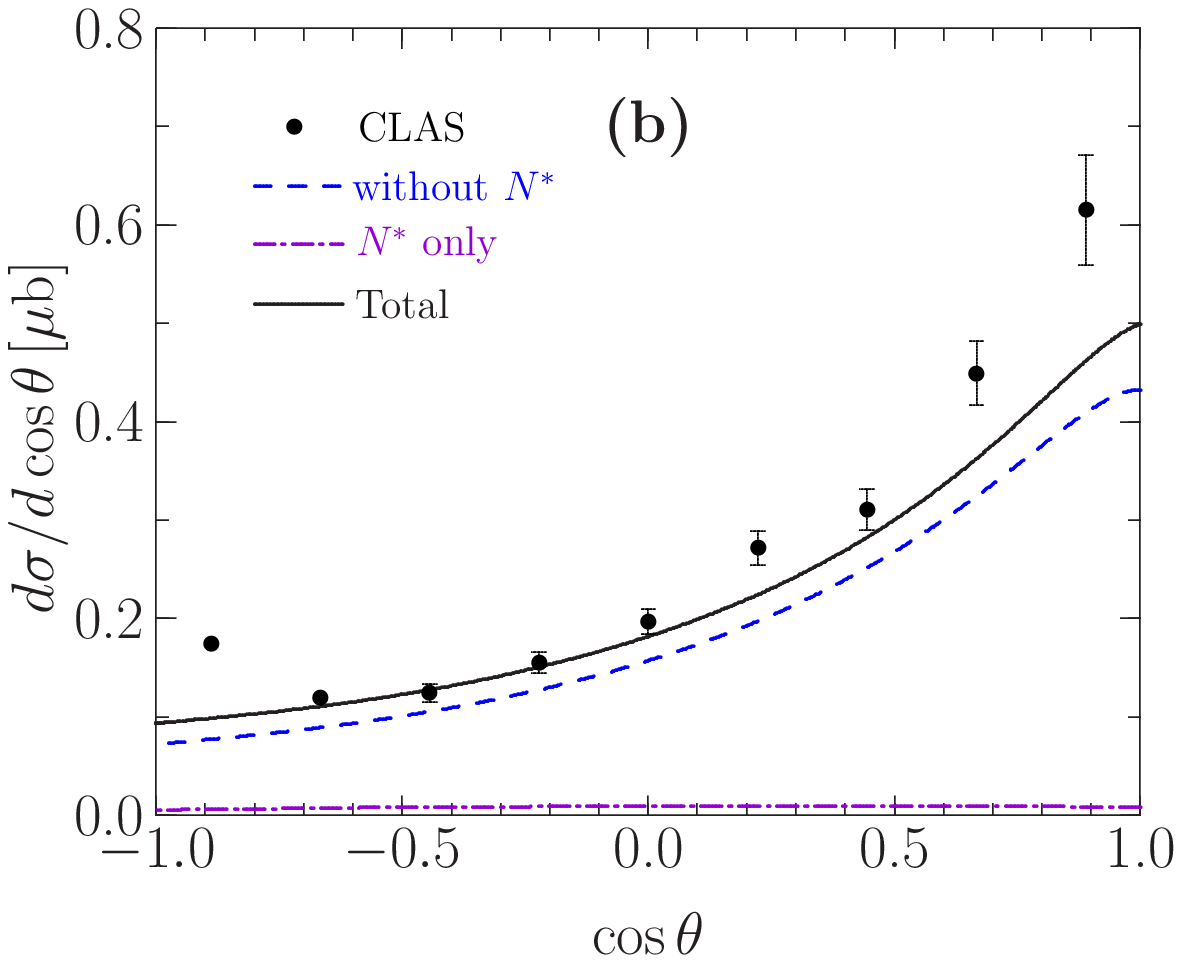}
\\ \bigskip \bigskip
\includegraphics[width=7.5cm]{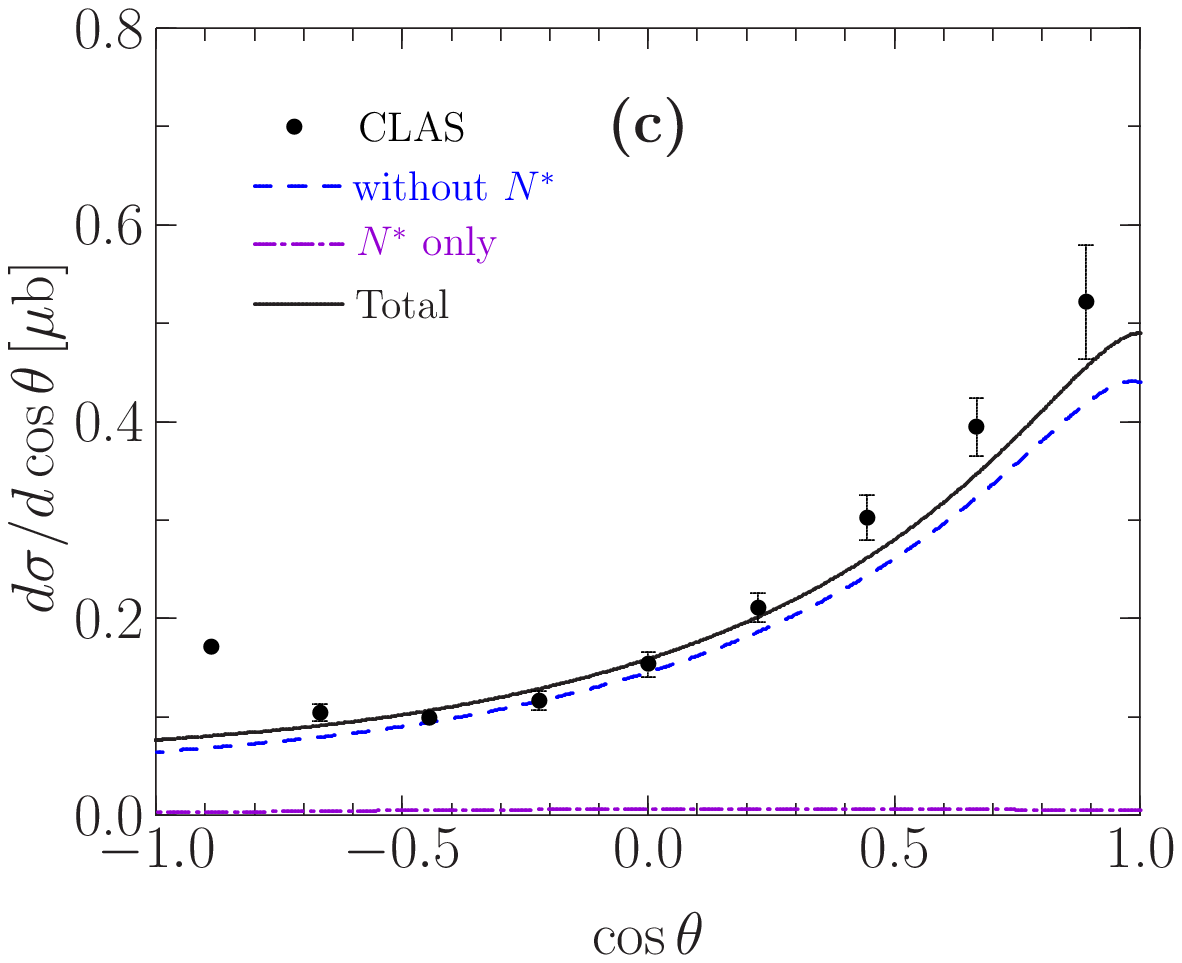} \qquad
\includegraphics[width=7.5cm]{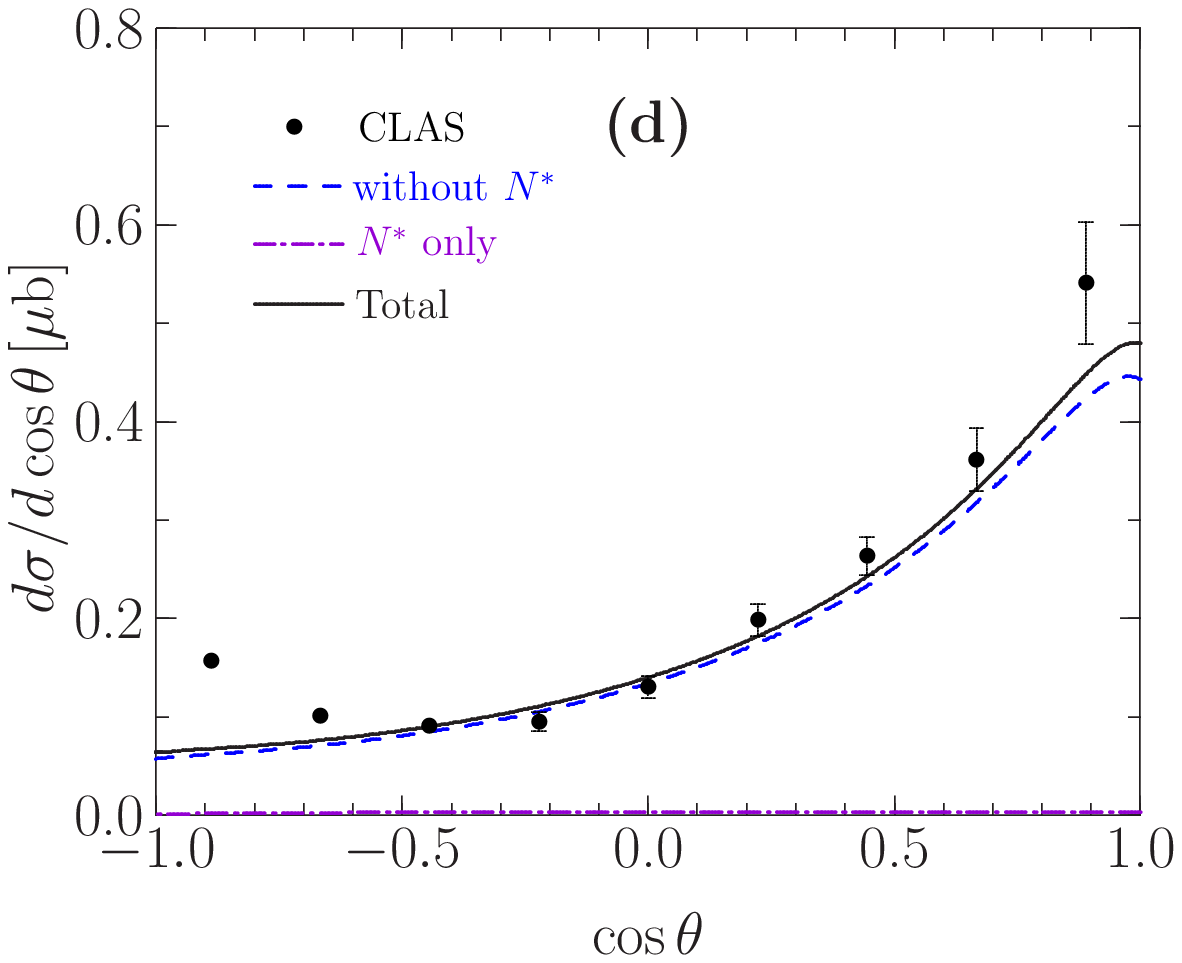}
\\ \bigskip \bigskip
\includegraphics[width=7.5cm]{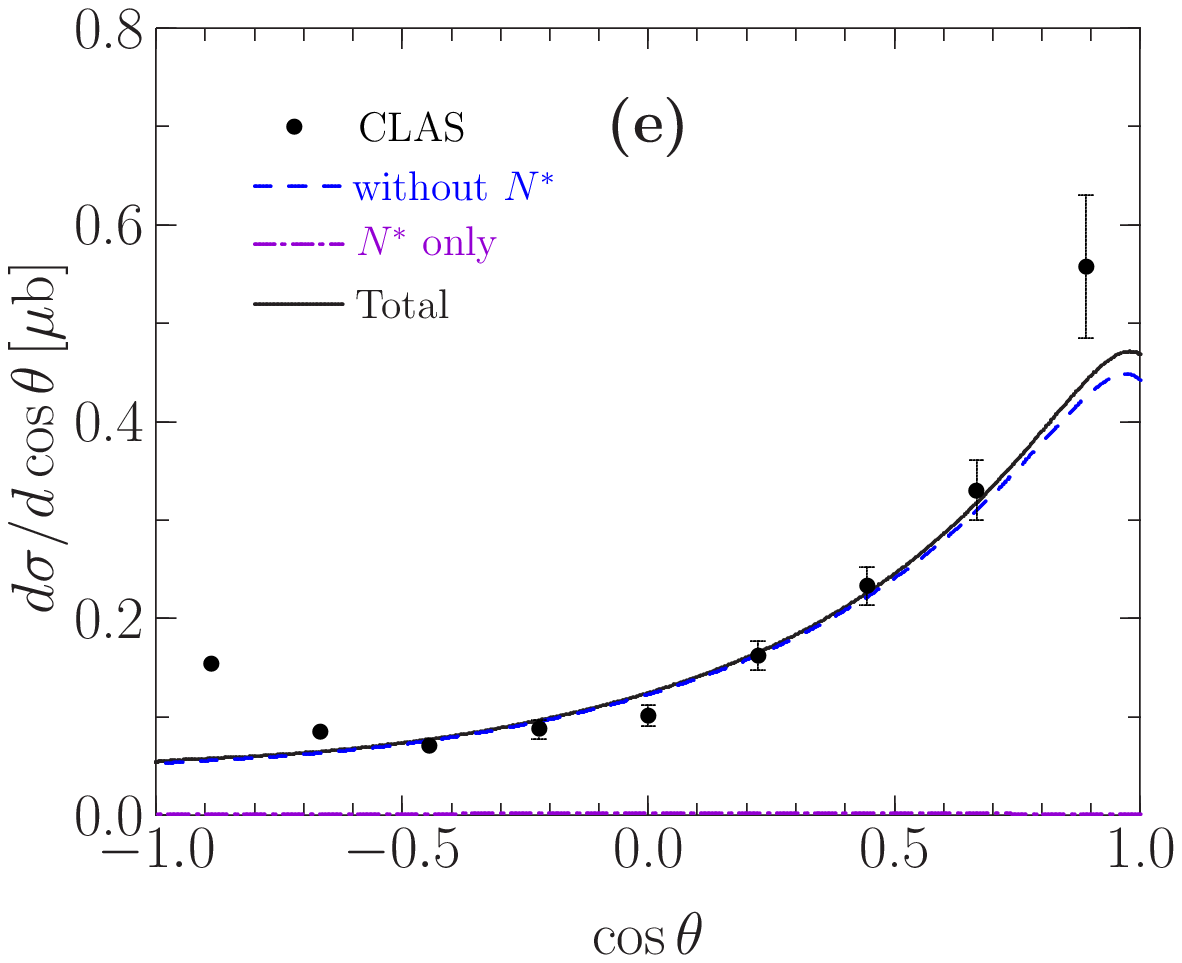} \qquad
\includegraphics[width=7.5cm]{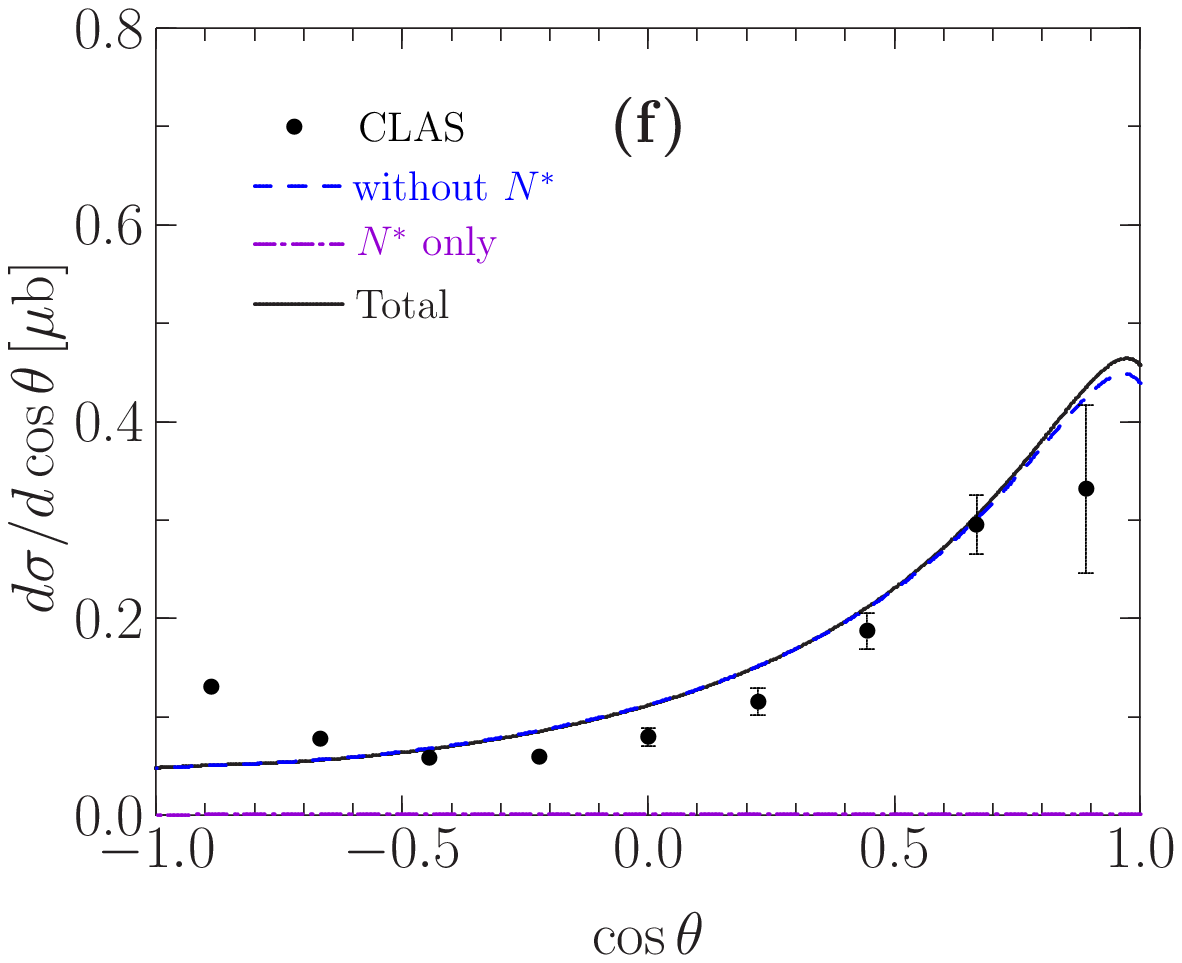}
\caption{ \label{FIG2}
(Color online) Differential cross sections $d\sigma/\cos\theta$ for
$\gamma p \to K^{*+} \Lambda$ at $E_\gamma = $ 
(a) $2.15$~GeV, (b) $2.25$~GeV, (c) $2.35$~GeV, (d) $2.45$~GeV,
(e) $2.55$~GeV, and (f) $2.65$~GeV. 
Dashed curves explain the background production mechanisms whereas 
dot-dashed ones are the contributions from nucleon resonances. The
solid curve draws the total contribution from all relevant diagrams.
The experimental data of the CLAS Collaboration are taken from
Ref.~\cite{HKT10}. }        
\end{figure}

In this Section, we present the results for various physical
observables of charged and neutral $K^*\Lambda$ photoproduction. Shown
in Fig.~\ref{FIG2} are the differential cross sections of $\gamma p
\to K^{*+}\Lambda$ for photon laboratory energies in the range of
$E_\gamma=2.15\,\mathrm{GeV} - 2.65\,\mathrm{GeV}$ as functions of
$\cos\theta$, where $\theta$ is the scattering angle of the $K^*$ in
the center-of-mass frame. Here, the dashed curves depict the cross
sections of the background production mechanisms and the contribution
from the considered nucleon resonances ($D_{13}$ and $D_{15}$) is
given by the dot-dashed curves. The solid ones draw their coherent
sums. The obtained results are compared with the experimental data of 
the CLAS collaboration at TJNAF~\cite{CLASKSTAR,HKT10}.  The present
results show that the cross sections are enhanced in the forward
direction because of the strong $K$-exchange contribution in the $t$
channel.  In the resonance region, i.e., $E_\gamma \approx 2$~GeV, the
effects of the nucleon resonances are obvious but become smaller as the
photon energy increases.  Although our model can generally reproduce the
cross section data qualitatively well, there is a discrepancy  
with the data in the backward scattering region.
It implies nontrivial contributions from hyperon resonances in
the $u$-channel at large scattering angles. The role of hyperon
resonances in $K^*$ photoproduction will be reported
elsewhere~\cite{FUTURE} (See, for example, Ref.~\cite{NY11}.) 

\begin{figure}[t]
\includegraphics[width=12cm]{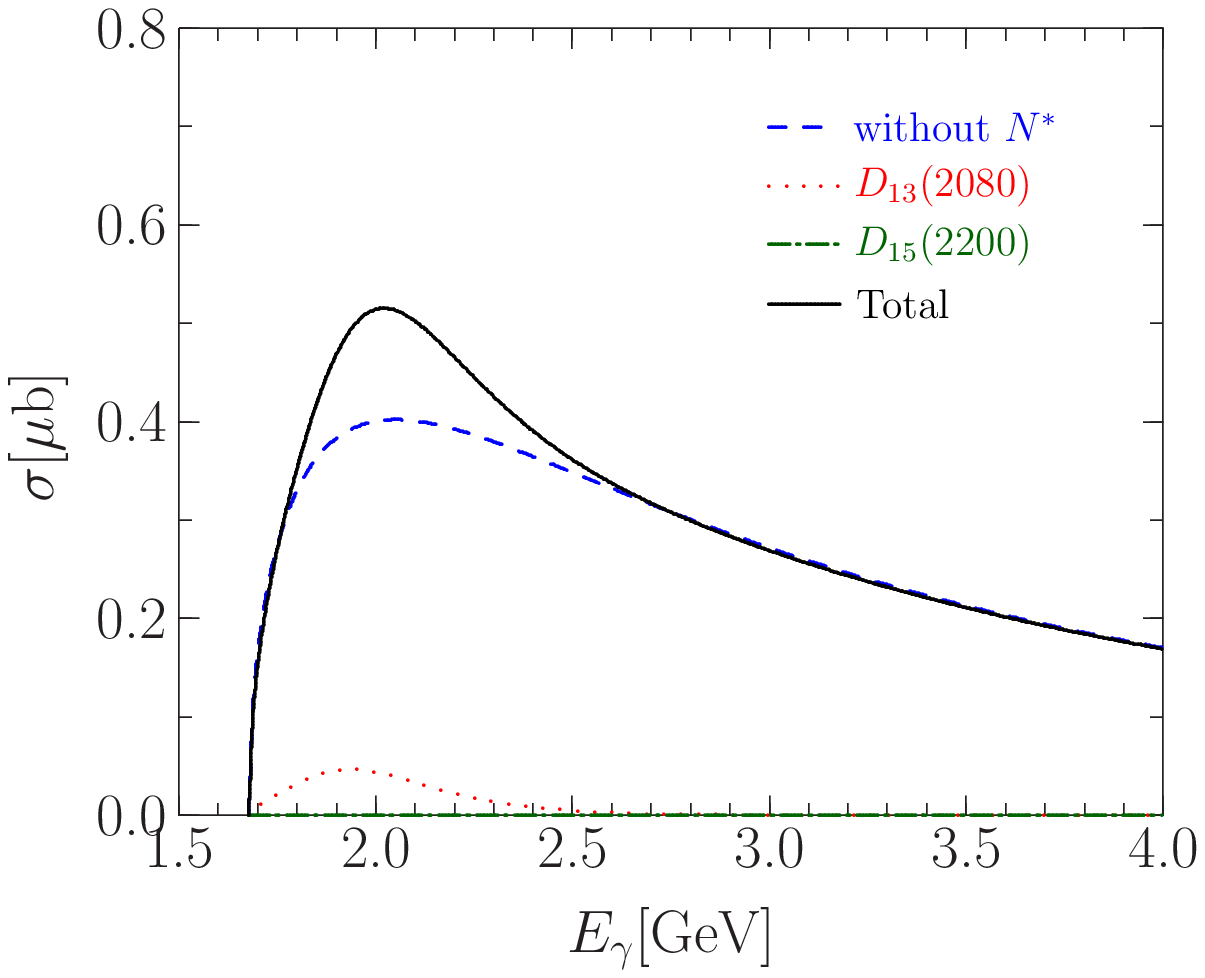} 
\caption{(Color online) Total cross sections for $\gamma p \to
K^{*+}\Lambda$. The contributions from the $D_{13}(2080)$ and the
$D_{15}(2200)$ are given by the dotted line and the dot-dash line,
respectively. The dashed line is the result without nucleon
resonances and the solid line is the full calculation.}        
\label{FIG3}
\end{figure}

In Fig.~\ref{FIG3}, we show the obtained total cross section for
$\gamma p \to K^{*+}\Lambda$ as a function of the photon laboratory
energy $E_\gamma$. Since Ref.~\cite{HKT10} does not report the total
cross section data, we do not compare  
our results with the data~\footnote{We have not included the preliminary
  data for the total cross sections for $\gamma p \to K^{*+}\Lambda$
  reported in Ref.~\cite{CLAS06e} because of an error committed in the
  analyses~\cite{FACTOR}.}. The results of the total cross section
exhibit a clear enhancement near
the threshold, It implies that higher resonances apearing in this
energy region play an essential role in explaining the mechanism of
the $K^*\Lambda$ photoproduction. In particular, the contribution from
the $D_{13}(2080)$ is prominent as shown by the
dotted curve, while that of the $D_{15}(2200)$ (the dot-dashed one) is
almost negligible. The dashed one depicts the background contribution,
which is dominated by the $K$-meson exchange in the $t$-channel. The 
contribution from the $\kappa$-meson exchange is small compared with
that from the $K$-meson exchange, although it is larger than the other
background diagrams near the threshold.  

\begin{figure}[t]
\mbox{}\bigskip
\includegraphics[width=7.5cm]{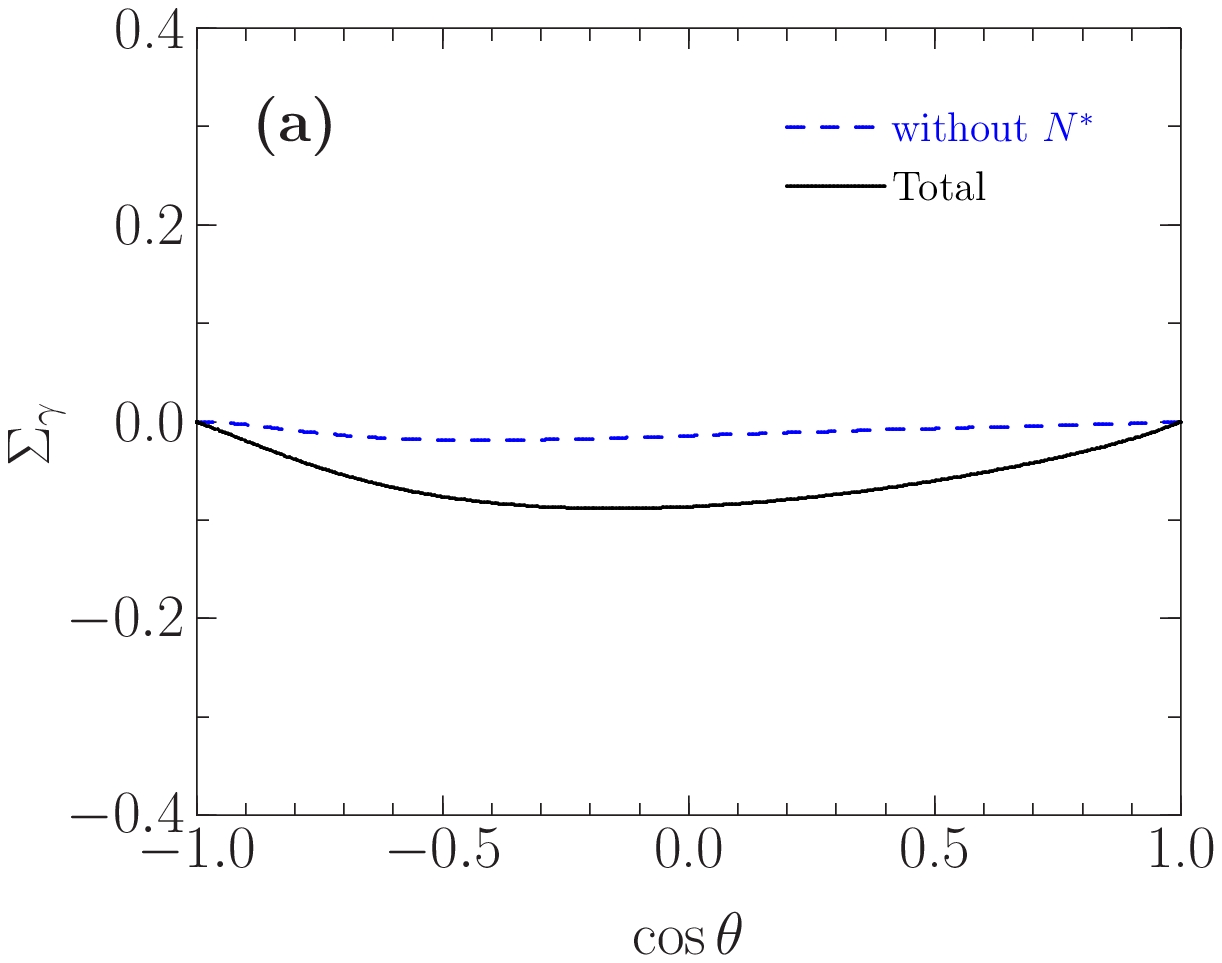} \qquad
\includegraphics[width=7.5cm]{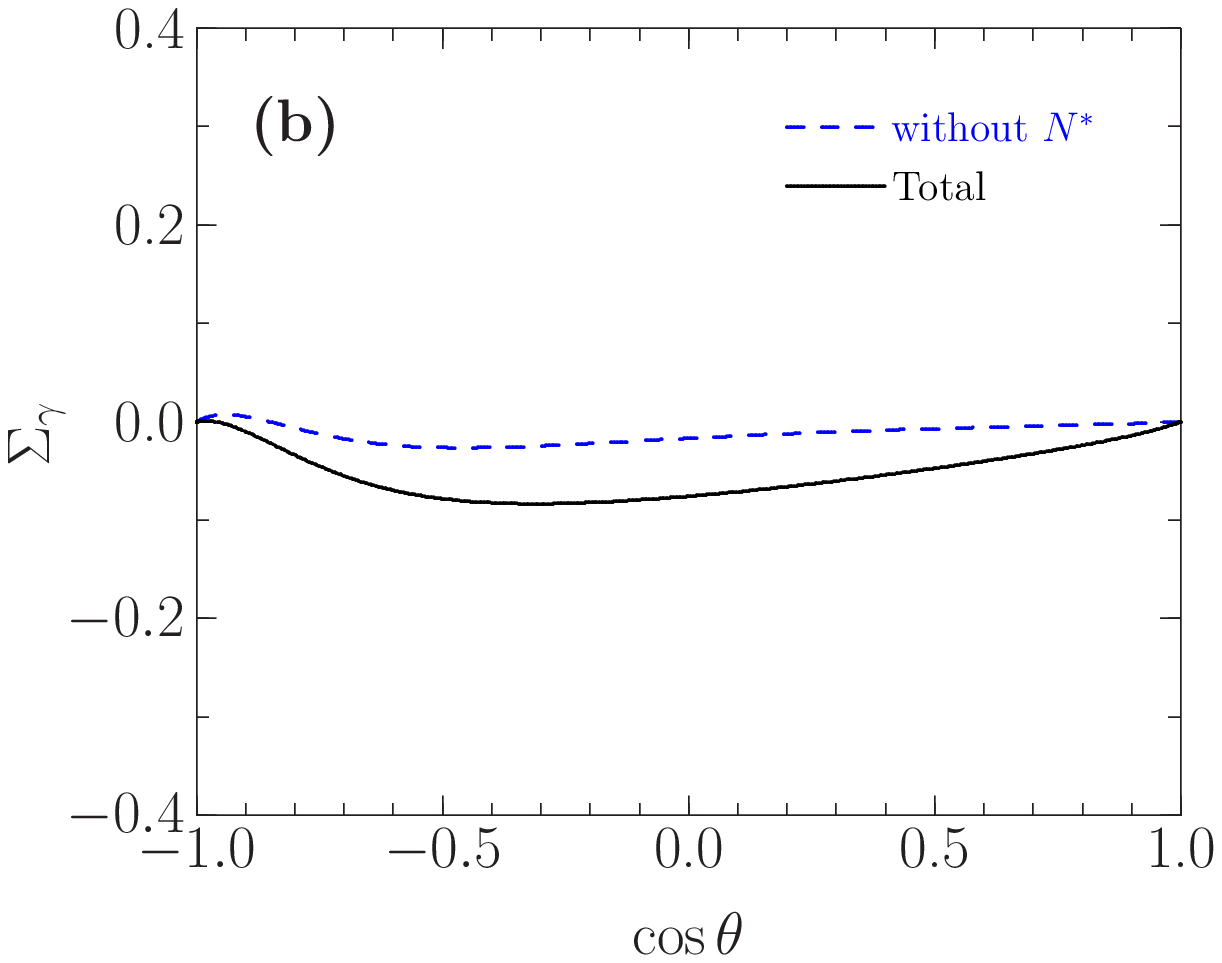}
\caption{(Color online) Photon-beam asymmetry $\Sigma_\gamma$ as a
  function of $\cos\theta$ for $\gamma p \to K^{*+}\Lambda$ at (a)
  $E_\gamma = 2.25$~GeV and (b) $E_\gamma = 2.55$~GeV.}        
\label{FIG5}
\bigskip
\end{figure}

\begin{figure}[t]

\includegraphics[width=7.5cm]{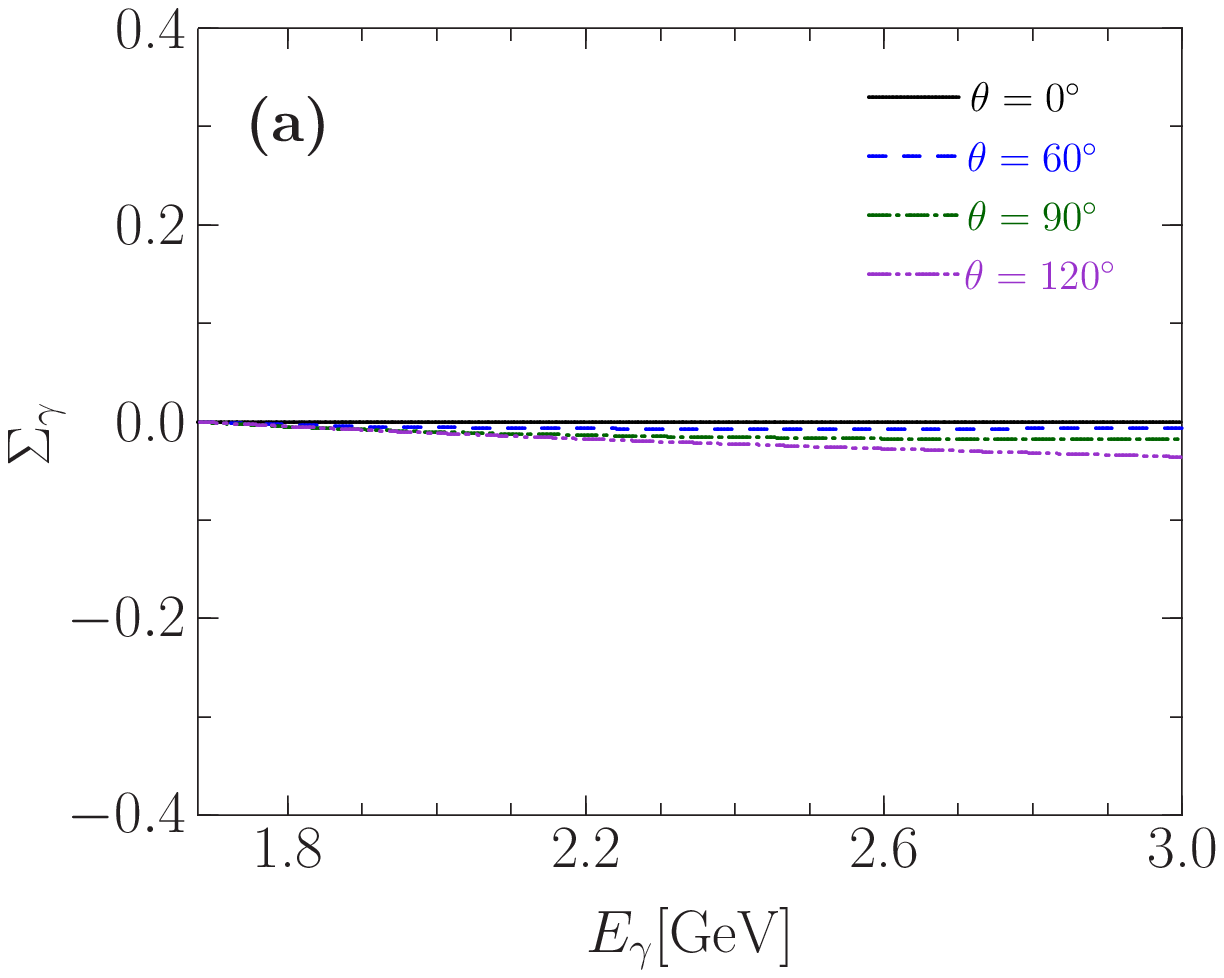} \qquad
\includegraphics[width=7.5cm]{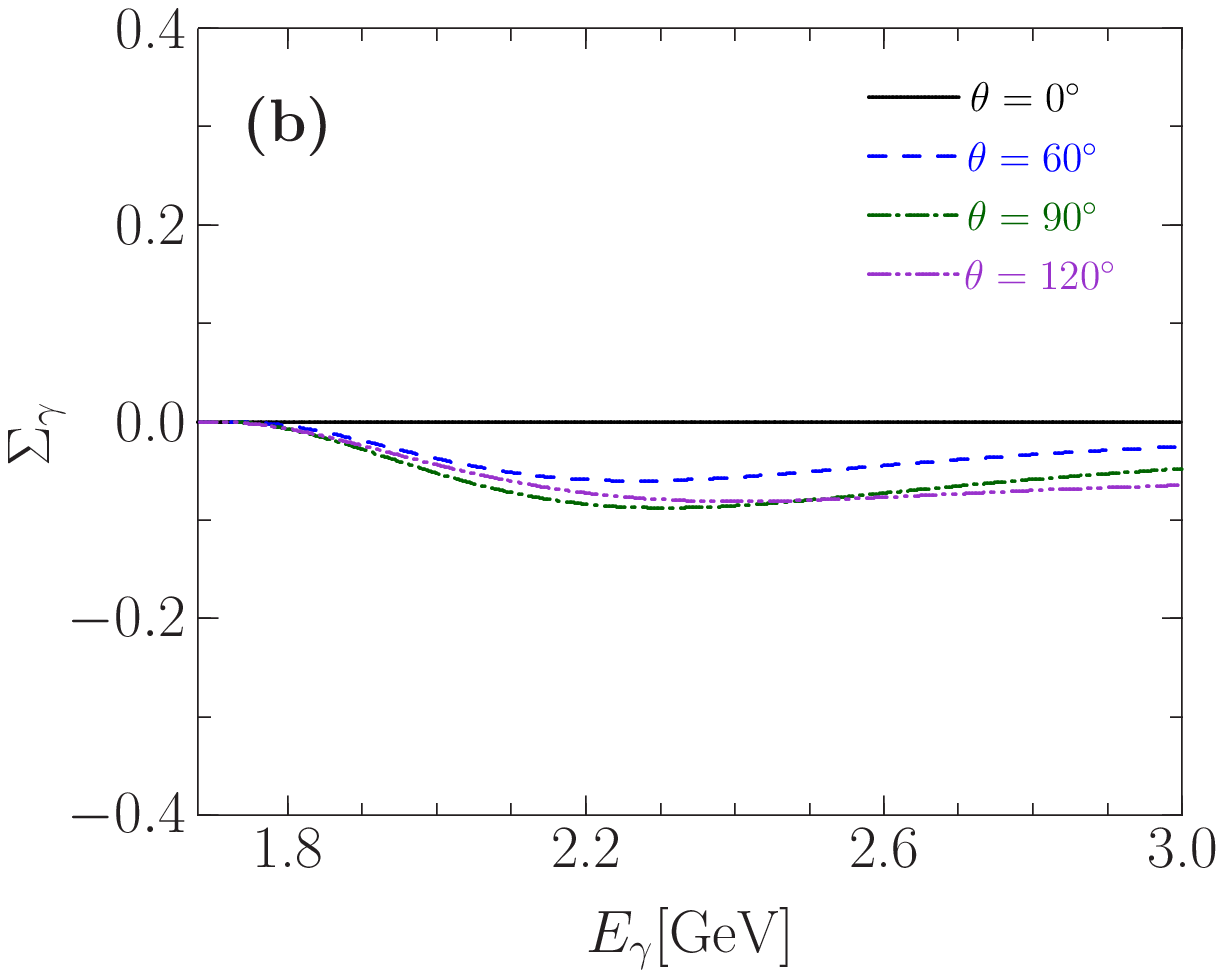}
\caption{(Color online) Photon-beam asymmetry $\Sigma_\gamma$ as a
  function of $E_\gamma$ for $\gamma p \to K^{*+}\Lambda$ at
  $\theta=0^\circ$, $60^\circ$, $90^\circ$, and $120^\circ$ (a)
  without the resonance contributions and (b) with the resonance
  contributions.}        
\label{FIG6}
\end{figure}

Using the same set of parameters, we can calculate the photon-beam 
asymmetry $\Sigma_\gamma$ that is defined as 
\begin{equation}
\label{eq:BA}
\Sigma_\gamma =
\frac{d\sigma_\parallel-d\sigma_\perp}{d\sigma_\parallel+d\sigma_\perp}, 
\end{equation}
where $\parallel$ ($\perp$) indicates that the direction of the
photon-polarization vector is parallel (perpendicular) to the reaction
plane which is defined by the outgoing $K^*$ momentum and the recoiled
$\Lambda$ momentum. In Fig.~\ref{FIG5}(a), the results of the
asymmetry $\Sigma_\gamma$ as a function of $\cos\theta$ at $E_\gamma =
2.25$~GeV are given and those at $E_\gamma=2.55$~GeV are drawn in
Fig.~\ref{FIG5}(b). The dashed curves are the results without the
nucleon resonances whereas the solid ones depict the
results with all contributions. We immediately find that the asymmetry
$\Sigma_\gamma \simeq 0$ without $N^*$ contributions. Because of
the strong magnetic-transition coupling of the nucleon resonances,
however, the beam asymmetry $\Sigma_\gamma$ turns negative with the
$N^*$ resonances included. Although the change is not drastically 
large, the size is noticeable enough to be tested by upcoming
experimental measurements, in particular, at $\theta \simeq
0^\circ$. This also can be seen in Fig.~\ref{FIG6}, where the
asymmetry $\Sigma_\gamma$ is given as a function of $E_\gamma$ at
scattering angle $\theta=0^\circ$, $60^\circ$, $90^\circ$, and
$120^\circ$, respectively. As shown in Fig.~\ref{FIG6}(a),
$\Sigma_\gamma$ is almost compatible with zero for the whole energy
region considered, though it increases mildly as $E_\gamma$
increases. On the other hand, as the photon energy increases, the beam
asymmetry becomes negative from the threshold and
its magnitude is getting increased up to around 2.2-2.4 GeV, depending
on the scattering angle. Then it slowly decreases. Thus, it indicates
that the $N^*$ resonances are involved in changing the polarization of
the $K^*$ from the original spin alignment of the photon beam. We 
anticipate that the future experimental data will further clarify this
point. 

\begin{figure}[t]
\bigskip
\includegraphics[width=12cm]{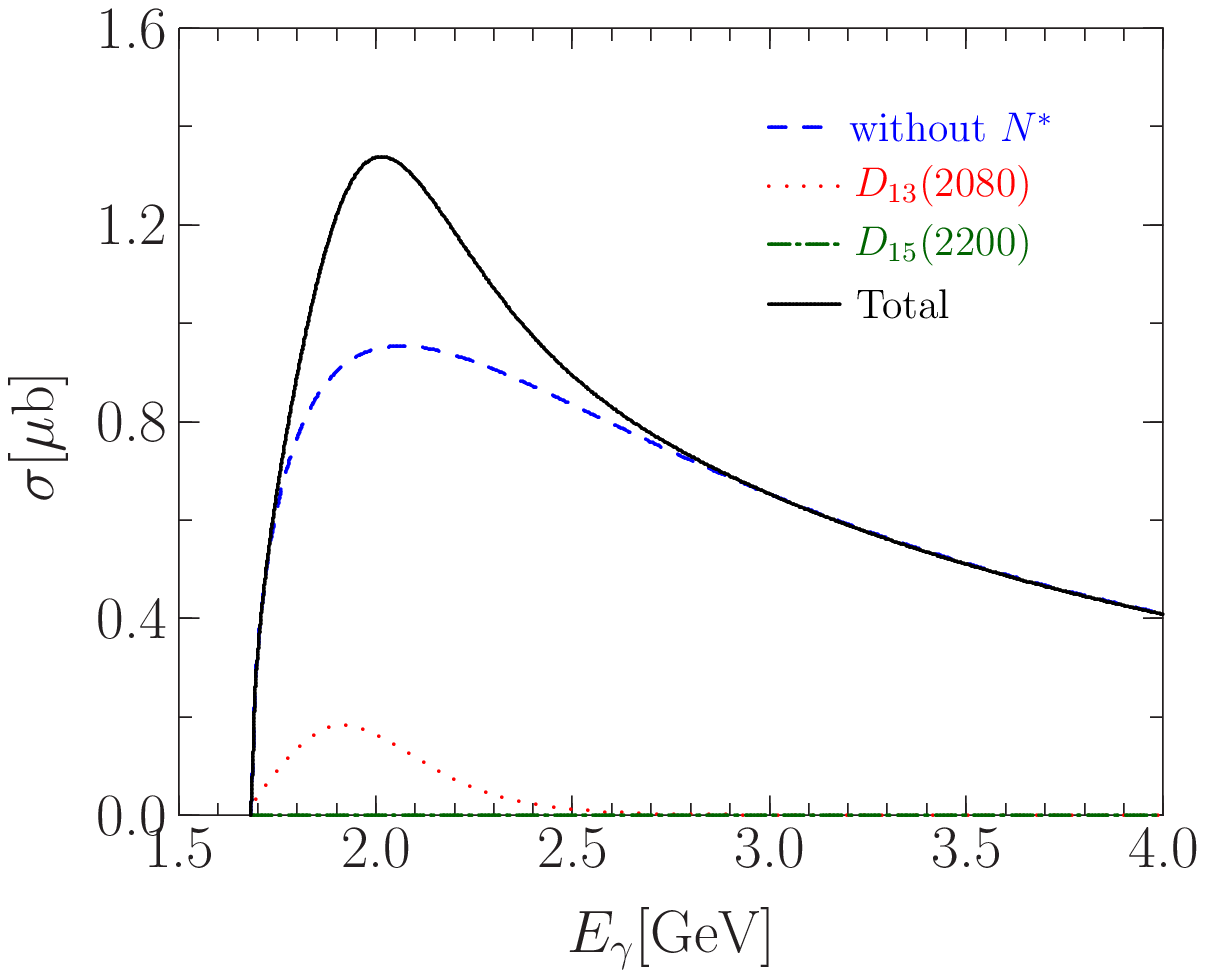}
\caption{(Color online) Total cross sections for $\gamma n \to
  K^{*0}\Lambda$. The contributions from the $D_{13}(2080)$ and the
  $D_{15}(2200)$ are given by the dotted line and the dot-dash line,
  respectively. The dashed line is the result without 
nucleon resonances and the solid line is the full calculation.}    
\label{FIG7}
\end{figure}

\begin{figure}[t]
\includegraphics[width=7.5cm]{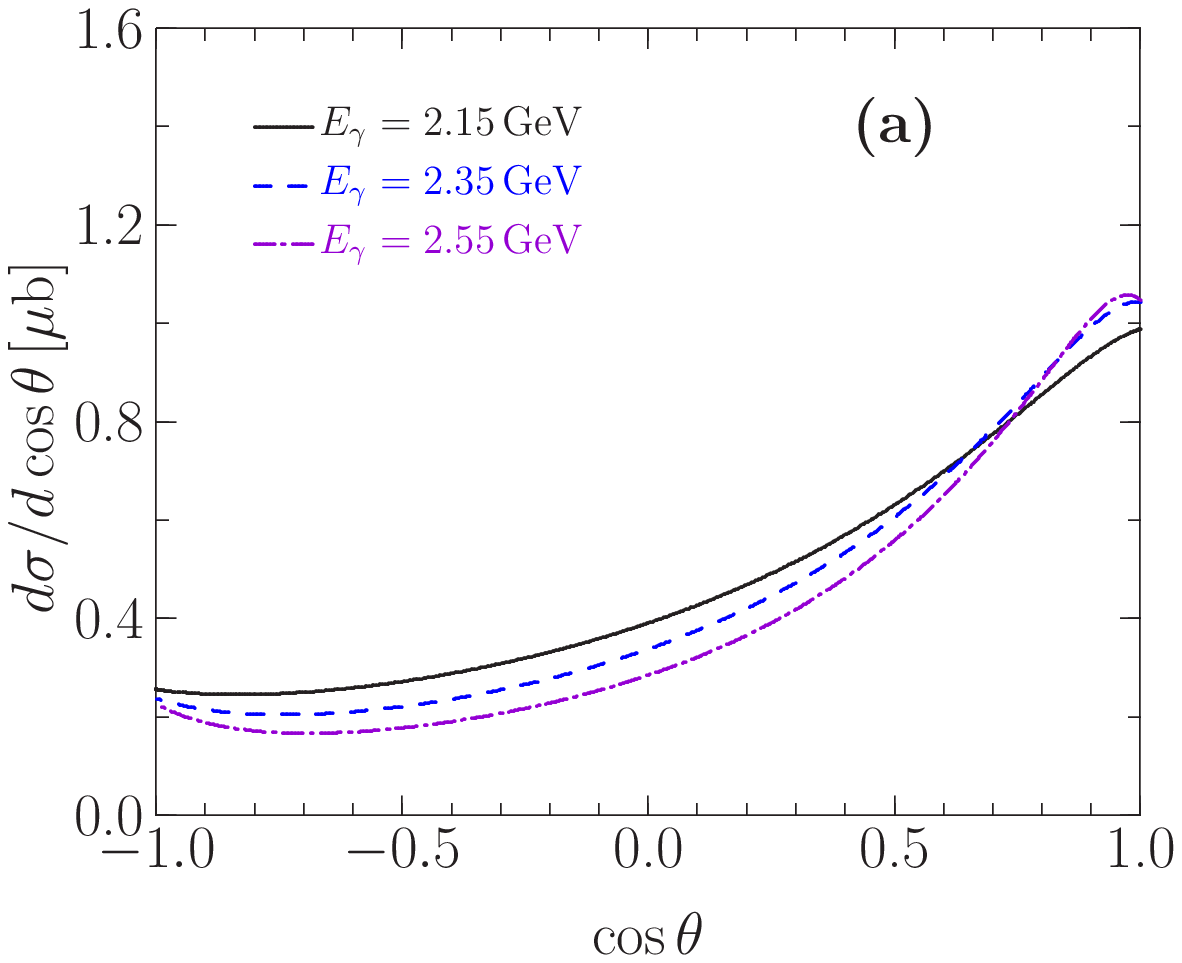} \qquad
\includegraphics[width=7.5cm]{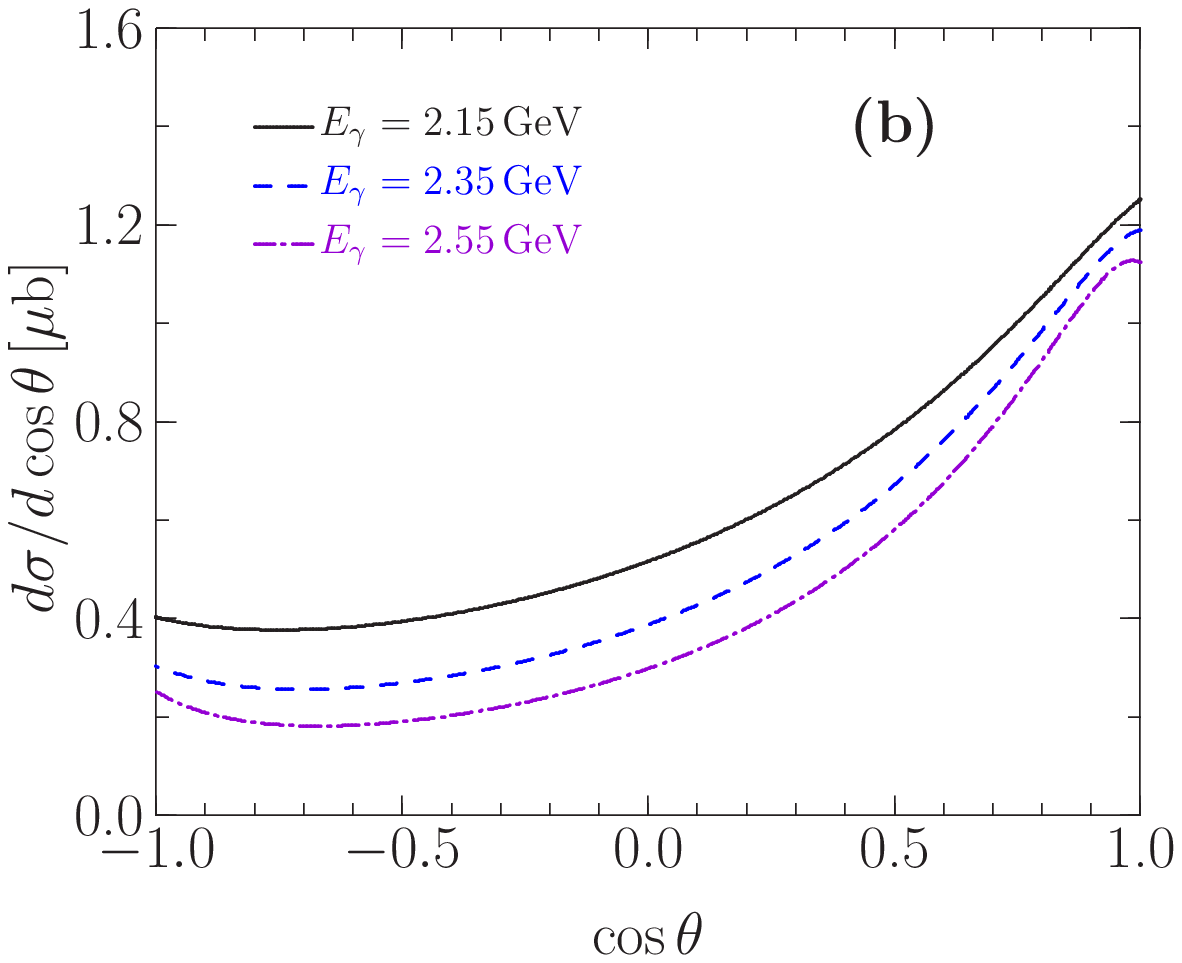}
\caption{(Color online) Differential cross section
  $d\sigma/d\cos\theta$  for $\gamma n \to K^{*0}\Lambda$  
at $E_\gamma=2.15$~GeV, $2.35$~GeV, and $2.55$~GeV (a) without $N^*$
contributions and (b) with $N^*$ contributions.}        
\label{FIG8}
\end{figure}

\begin{figure}[t]
\bigskip
\includegraphics[width=7.5cm]{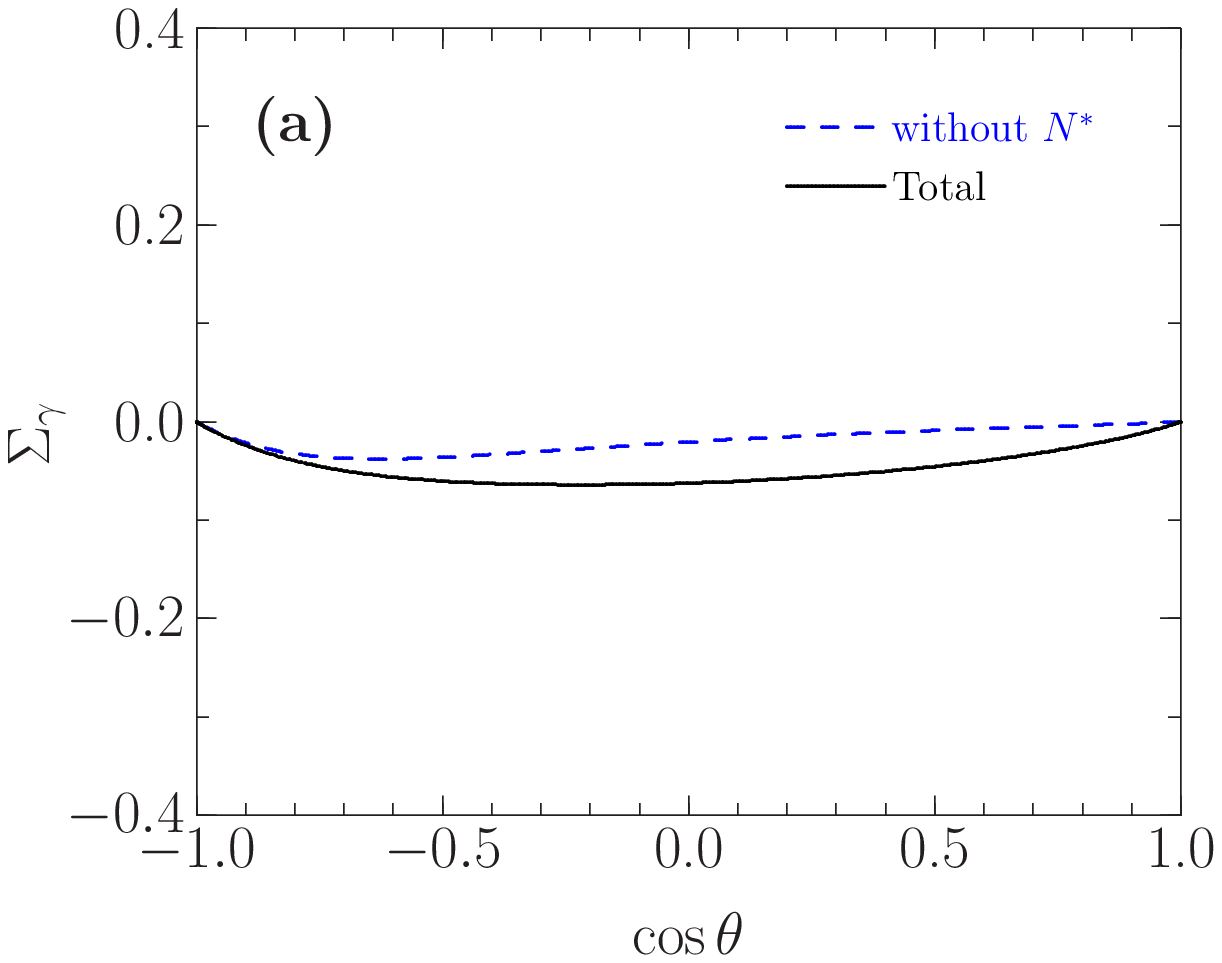} \qquad
\includegraphics[width=7.5cm]{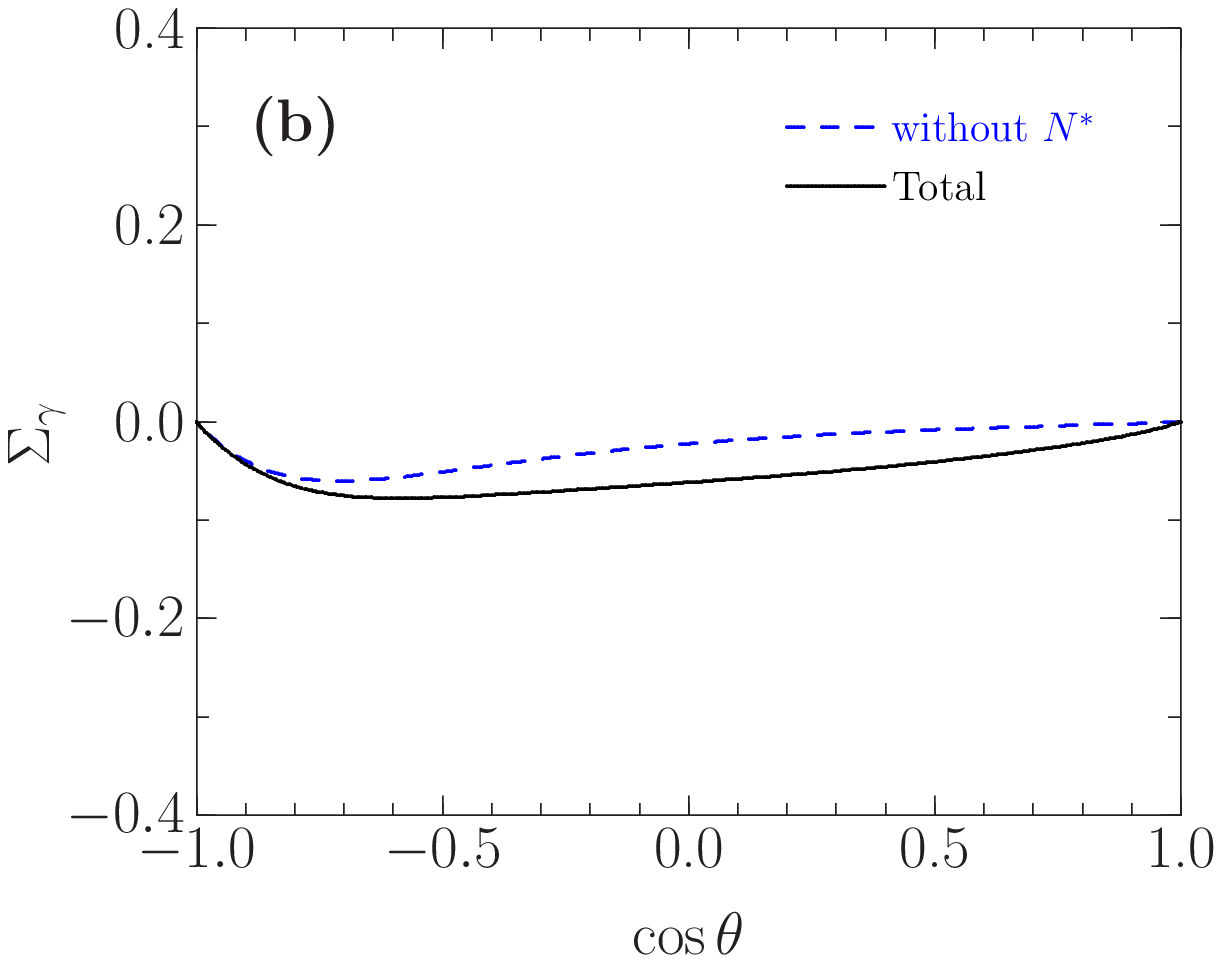}
\caption{(Color online) Photon-beam asymmetry $\Sigma_\gamma$ for
  $\gamma n \to K^{*0}\Lambda$ at (a) $E_\gamma=2.25$~GeV and (b)
  $E_\gamma = 2.55$~GeV.}        
\label{FIG10}
\end{figure}

\begin{figure}[t]
\bigskip\bigskip
\includegraphics[width=7.5cm]{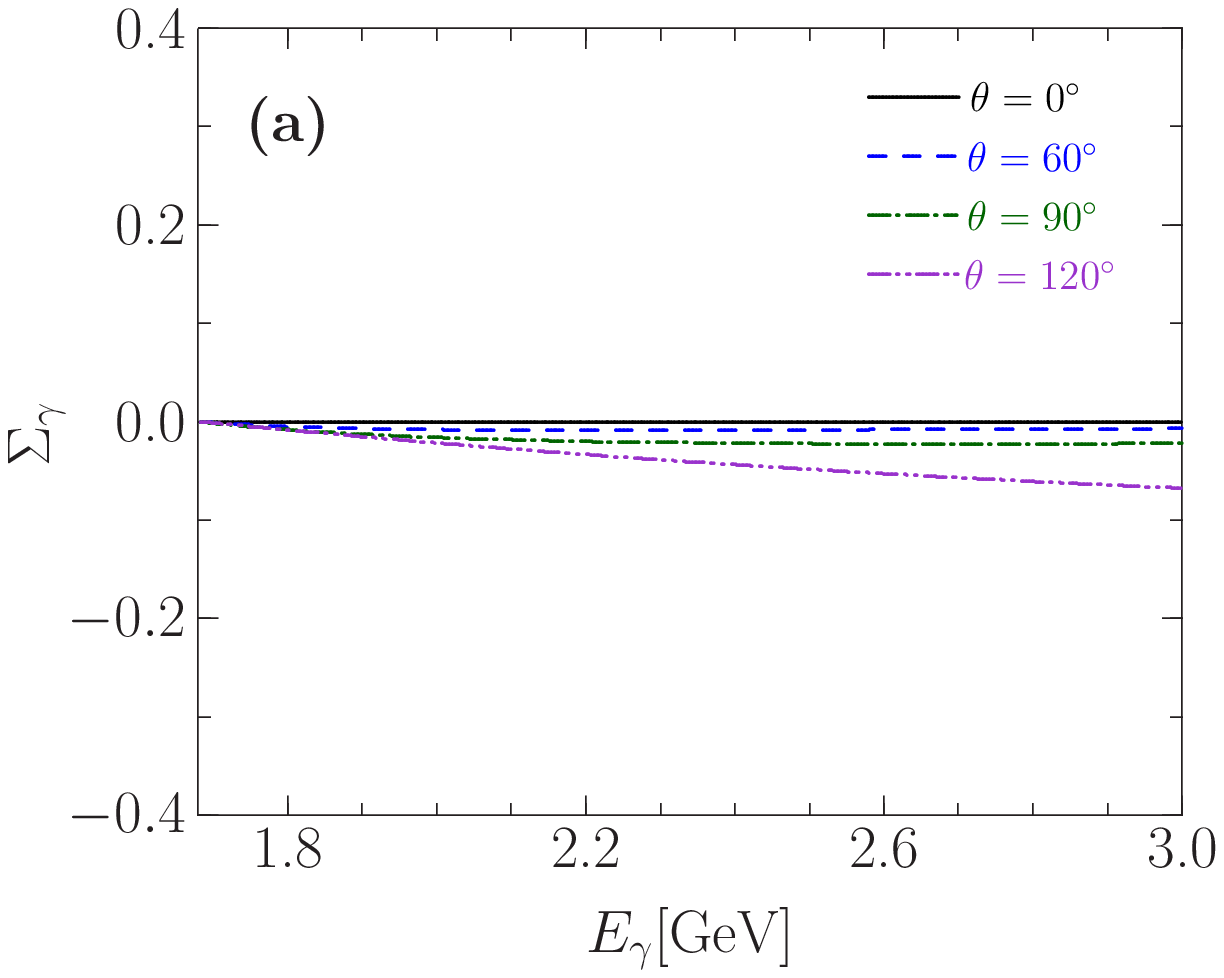} \qquad
\includegraphics[width=7.5cm]{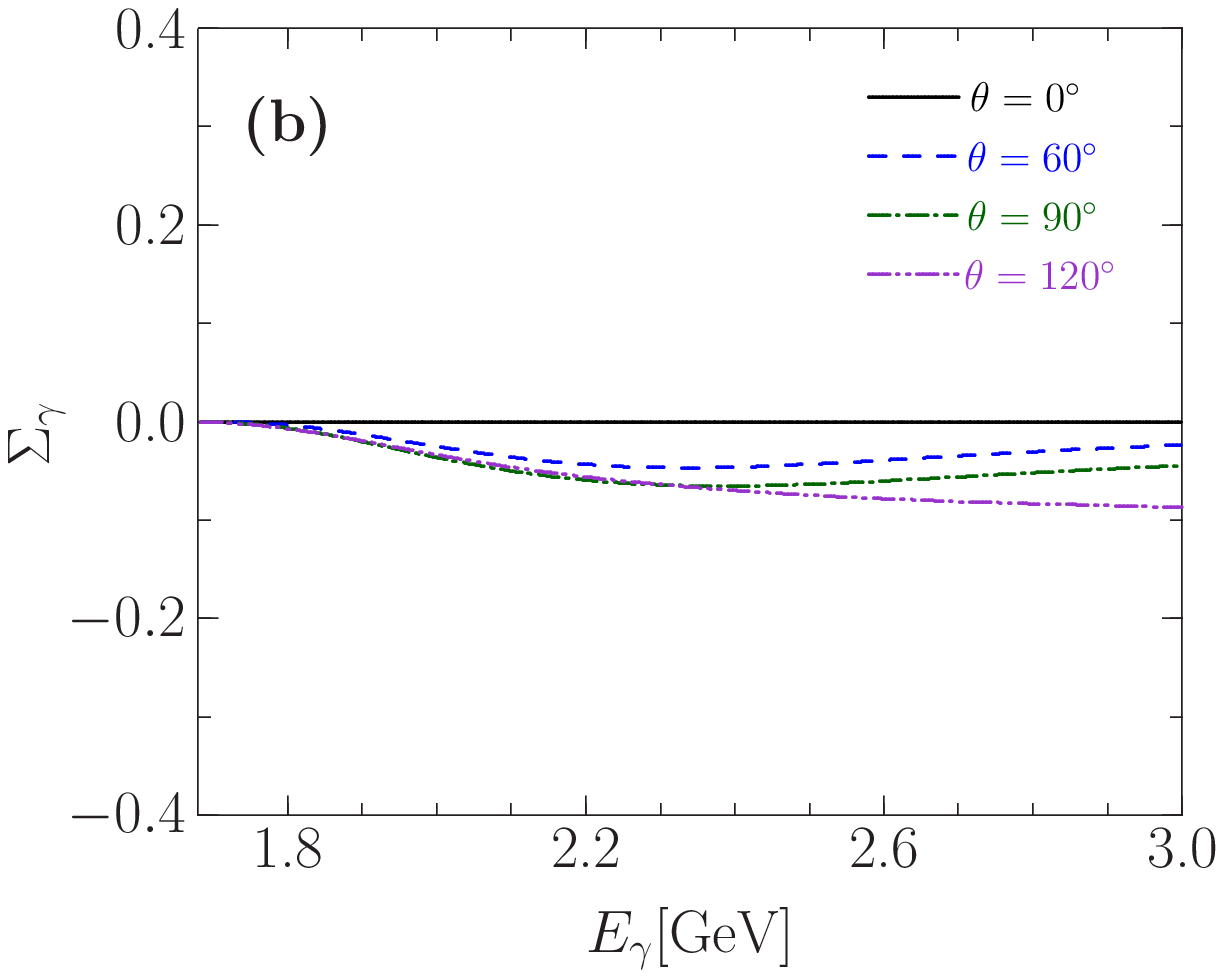}
\caption{(Color online) Photon-beam asymmetry $\Sigma_\gamma$ as a
  function of $E_\gamma$ for $\gamma n \to K^{*0}\Lambda$ at
  $\theta=0^\circ$, $60^\circ$, $90^\circ$, and $120^\circ$ (a)
  without the resonance contributions and (b) with the resonance
  contributions.}         
\label{FIG11}
\end{figure}

Figures~\ref{FIG7}--\ref{FIG11} predict the observables of the $\gamma
n \to K^{*0} \Lambda$ reaction. Shown in Fig.~\ref{FIG7} is the total
cross section for this reaction as a function of $E_\gamma$. Because
of the neutral charge of the $K^{*0}$, the $K^*$-exchange and the
contact term are absent in this reaction as well as the electric
photon-hyperon coupling. As in $\gamma p \to K^{*+}\Lambda$, the
background production mechanisms are dominated by the $K$-meson
exchange. However, since the neutral coupling of the $\gamma NN^*$
interaction is larger than that of the charged coupling by a factor of
about $\sqrt2$, the total cross section for $\gamma n \to
K^{*0}\Lambda$ turns out to be larger than those of $\gamma p \to
K^{*+}\Lambda$ by a factor of 2. As shown in Fig.~\ref{FIG7}, we again
have considerable contributions from the $D_{13}$ resonance near the
threshold, whereas that from the $D_{15}$ 
remains negligible.  

The differential cross sections for $\gamma n \to K^{*0}\Lambda$ are
shown in Fig.~\ref{FIG8} at $E_\gamma = 2.15$~GeV, $2.35$~GeV, and
$2.55$~GeV, and the corresponding results are similar to   
those of $\gamma p \to K^{*+}\Lambda$. Unlike the charged $K^*$
photoproduction, the differential cross sections slightly increase at
the very backward scattering angle. We find that this behavior is
caused by the interference among the $u$-channel contributions which 
depends on the isospin of the targets.

The results of the asymmetry $\Sigma_\gamma$ are presented in 
Figs.~\ref{FIG10} and \ref{FIG11}. Figure~\ref{FIG10} draws the 
angular distribution of $\Sigma_\gamma$ at $E_\gamma = 2.25$~GeV and
$2.55$~GeV, whereas Fig.~\ref{FIG11} shows the energy dependence of 
$\Sigma_\gamma$. The results are similar to thse of $\gamma p \to
K^{*+}\Lambda$. Namely, the contributions from the nucleon resonances 
change the asymmetry $\Sigma_\gamma$ as in the case of the proton 
target.    

\section{Summary and conclusion}
In this work, we have studied $K^*\Lambda$ photoproduction off the
nucleon target, i.e. $\gamma N\to K^*\Lambda$, focusing on the role of
nucleon resonances, in particular, $D_{13}(2080)$ and $D_{15}(2200)$,
the masses of which are close to the threshold of the reaction. 
We found that the recent differential cross section data of the CLAS
Collaboration for the reaction of $\gamma p \to K^{*+}\Lambda$ were
be reasonably well described by the present model except for the
backward scattering region. It indicates that further
investigations on the role of $u$-channel hyperon resonances are
required~\cite{FUTURE}. The contributions from the nucleon resonances
were found to be dominated by the $D_{13}(2080)$ and the $D_{15}$
provides almost no corrections. The resonance parameters of the
$D_{13}(2080)$ are determined by the experimental 
data of PDG~\cite{PDG10} and the theoretical estimate of
Ref.~\cite{Caps92,CR98b}. Although it requires more sophisticated
analyses to precisely constrain the resonance parameters of the
$D_{13}(2080)$, the present investigation shows that nucleon
resonances contribute nontrivially to $K^*\Lambda$
photoproduction. It was also shown that the beam asymmetry was changed
noticeably by the $N^*$ resonances. The upcoming experimental data
from the CLAS collabotation or from the LEPS collaboration will judge
the role of these $N^*$ resonances in $K^*$ photoproduction.

The present model was also applied to the reaction of $\gamma n \to 
K^{*0}\Lambda$. We predicted the cross sections and the beam asymmetry of
this reaction. We confirm that this reaction has larger cross sections
than the $\gamma p \to K^{*+}\Lambda$ reaction, and the role of
nucleon resonances is played in $\gamma n \to K^{*0}\Lambda$ as
significantly as in $\gamma p \to K^{*+}\Lambda$. We expect that
higher excited resonances of the nucleon may shed light on 
other photoproduction processes such as $\gamma
N \to K^* \Sigma$. The corresponding investigation is under way.

\section*{Acknowledgment}
We are grateful to K.~Hicks for kindly providing us with the
experimental data and for constructive comments. We also thank
A.~Hosaka and C.~W.~Kao for fruitful discussions. Two of us
(S.H.K. and H.Ch.K.) were supported by Basic Science Research Program
through the National Research Foundation of Korea funded by the
Ministry of Education, Science and Technology (Grant Number:
2009-0089525). The work of S.i.N. was partially supported by the
Grant NRF-2010-0013279 from the National Research Foundation of
Korea. Y.O. was supported by the National Research Foundation of Korea
Grant funded by the Korean Government (NRF-2011-220-C00011).

\end{document}